\documentclass[preprint]{aastex}
\usepackage{mathrsfs}
\usepackage{graphicx}
\usepackage{threeparttable}
\usepackage{multirow}
\usepackage{colortbl}
\definecolor{mypink}{rgb}{.99,.91,.95}
\usepackage{booktabs}
\usepackage{color}

\begin{document}

\title{A Mid-infrared Flare in the Active Galaxy MCG-02-04-026: Dust Echo of a Nuclear Transient Event}

\author{Luming Sun\altaffilmark{1,2}, Ning Jiang\altaffilmark{1,2}, Tinggui Wang\altaffilmark{1,2}, Hongyan Zhou\altaffilmark{3,1,2}, Liming Dou\altaffilmark{4}, Chenwei Yang\altaffilmark{3}, Xiang Pan\altaffilmark{3}, Zhenfeng Sheng\altaffilmark{1,2}, Zhihao Zhong\altaffilmark{1,2}, Lin Yan\altaffilmark{5}, Ge Li\altaffilmark{1}}
\altaffiltext{1}{Key laboratory for Research in Galaxies and Cosmology, Department of Astronomy, University of Science and Technology of China, Chinese Academy of Sciences, Hefei, Anhui 230026, China, lmsun@ustc.edu.cn, jnac@ustc.edu.cn}
\altaffiltext{2}{School of Astronomy and Space Science, University of Science and Technology of China, Hefei 230026, China}
\altaffiltext{3}{Polar Research Institute of China, 451 Jinqiao Road, Shanghai, China}
\altaffiltext{4}{Center for Astrophysics, Guangzhou University, Guangzhou, 510006, China}
\altaffiltext{5}{Caltech Optical Observatories, Cahill Center for Astronomy and Astrophysics, California Institute of Technology, Pasadena,
CA 91125, USA}

\begin{abstract}
  We report the discovery of a mid-infrared (MIR) flare using WISE data in the center of the nearby Seyfert 1.9 galaxy MCG-02-04-026.
  The MIR flare began in the first half of 2014, peaked around the end of 2015, and faded in 2017.
  During these years, energy more than $7\times10^{50}$ erg was released in the infrared, and the flare's MIR color was generally turning red.
  We detected neither optical nor ultraviolet (UV) variation corresponding to the MIR flare based on available data.
  We explained the MIR flare using a dust echo model in which the radiative transfer is involved.
  The MIR flare can be well explained as thermal reradiation from dust heated by UV-optical photons of a primary nuclear transient event.
  Although the transient event was not seen directly due to dust obscuration, we can infer that it may produce a total energy of at least $\sim10^{51}$ erg, most of which was released in less than $\sim$3 years.
  The nature of the transient event could be a stellar tidal disruption event by the central supermassive black hole (SMBH), or a sudden enhancement of the existing accretion flow onto the SMBH, or a supernova which was particularly bright.
\end{abstract}

\section{Introduction}

Recently, in the nuclear regions of some active galaxies, transient events involving rapid and great rise in the luminosity have been reported.
The examples are CSS100217:102913+404220 (Drake et al. 2011), PS16dtm (Blanchard et al. 2017), an event in F01004-2237 (Tadhunter et al. 2017), PS1-10adi (Kankare et al. 2017), an event in W0948+0318 (Kankare et al. 2017; Assef et al. 2018), AT 2017bgt (Trakhtenbrot et al. 2019), OGLE17aaj (Gromadzki et al. 2019), and some more (see Lawrence et al. 2016; Graham et al. 2017; Kankare et al. 2017; Kostrzewa-Rutkowska et al. 2018; Drake et al. 2019; etc.).
During these events, the galactic nuclei reached their peak luminosities in one to two months with their magnitudes decreasing by $\sim$1 magnitudes or even more, out of the common range of active galactic nuclei (AGN) variabilities (generally less than 0.5 magnitude, Caplar et al. 2017).
The nuclei then gradually faded, and their luminosities returned to the initial levels in a half to a couple of years.
These events evolved much faster than another type of AGN transient events with time scales of ten years or longer, such as Mrk 1018 (McElroy et al. 2016) and Mrk 590 (Denney et al. 2014), which were called ``changing look AGN'' due to the changes in their spectral types.

The natures of these transient events are not entirely understood.
Researchers have suggested three possible origins for these events, including stellar tidal disruption events (TDE) by the central supermassive black hole (SMBH) (e.g., Blanchard et al. 2017; Tadhunter et al. 2017), super-luminous supernova (SLSNe) (e.g., Drake et al. 2011), and an enhancement of the existing accretion flow of the AGN (e.g., Trakhtenbrot et al. 2019; Gromadzki et al. 2019).
TDEs and SLSNe are rare and bright transient events that were first recognized in inactive galaxies.
A TDE occurs when a star passes by a SMBH and is tidally disrupted (e.g., Rees 1988).
The disrupted stellar materials are then accreted by the SMBH, generating a luminous flare.
The flare's luminosity peaks in months or shorter, and then decreases over a period of months to years (Komossa 2015).
The TDE flare can cause the inactive galaxy that it resides in to temporarily look like an active galaxy.
SLSNe (e.g., Gal-Yam 2012) are a rare class of supernovae (SNe) that are particularly bright (less than $-21$ in magnitude in optical, Moriya 2018).
Their luminosities, duration time, and the shapes of their light curves (LC) are similar to those of TDEs.
The nature of SLSNe and the reason why they are so bright are still unclear.
According to their spectra, SLSNe can be divided into two types.
One type is SLSNe-I with no Hydrogen lines, and the other type is SLSNe-II with Hydrogen lines, and the spectral behaviors of some SLSNe-II can be similar to those of TDEs and AGN.
The time scale, energy budget and spectral features of most of the transient events in active galaxies mentioned in the first paragraph are similar to those of TDEs and SLSNe-II in inactive galaxies, thus they may be these two types of events that occur in active galaxies.
Sometimes, the LC or spectral features of some transient events are difficult to be explained with TDE or SLSNe-II, and researchers tended to interpret them as sudden enhancements of the existing accretion flow onto the SMBH (e.g., Trakhtenbrot et al. 2019), though the reason why the accretion rate can change drastically in such a short time is not clear (see Lawrance 2018).
Because the phenomena produced by these three possible origins are similar, despite great efforts of researchers, no unambiguous conclusions can be reached on the nature of the transient events in active galaxies.

Many of the transient events in active galaxies are accompanied by mid-infrared (MIR) flares.
The examples are as CSS100217:102913+404220 (Jiang et al. 2019), PS16dtm (Jiang et al. 2017), F01004-2237 (Dou et al. 2017), PS1-10adi (Kankare et al. 2017), W0948+0318 (Assef et al. 2018), OGLE17aaj (Yang et al. 2019).
The MIR flares were observed by an all-sky time-domain survey in the MIR band named {\it Wide field Infrared Survey Explorer} (WISE), which we will introduce in detail later.
The MIR flares generally lasted for several to ten years, during which they dramatically strengthened the MIR radiation of the galaxies as the magnitudes of the galaxies decreased by a half to several magnitudes.

On the origin of the accompanied MIR flares in the transient events, a promising interpretation is that the MIR flares come from dust echoes of the transient events.
The dust echo has been modeled by Lu et al. (2016) and Jiang et al. (2017).
Briefly, when a transient event occurs in a galactic nucleus, its growing UV-optical radiation sublimates the dust in the environment continuously, forcing the dust to form inner surfaces with increasing radii.
When the UV-optical radiation declines, it can no longer sublimate the dust, and hence the radius of the inner surface stops increasing.
The UV-optical radiation heats the dust around the eventual inner surface, and meanwhile is converted to IR reradiation by the dust.
The model is supported by increasing observational evidence.
The MIR radiation peaked later than the UV-optical radiation, and there is a rough relationship between the time delay $\tau$ between UV-optical and MIR radiations and the sublimation radius ($r_{\rm sub}$) corresponding to the peak luminosity ($L_{\rm max}$) of the transient event: $\tau \sim R_{\rm sub}(L_{\rm max}) / c$ (e.g., Jiang et al. 2017; Dou et al. 2017).
The relationship is the observational evidence that the echoing dust is located around the eventual inner surface.
Blanchard et al. (2017) found that PS16dtm, a transient event hosted in a narrow line Seyfert 1 galaxy with existing Fe II emission, enhanced the Fe II emission in half a year after the transient event's flux peaks.
Also, Jiang et al. (2019) noticed that along with PS1-10adi, transient Fe II emission increased rapidly around the time of the optical maximum and disappeared in an observation 3.5 years later.
These transient Fe II emissions that varied on a time scale of years were interpreted as the radiation of Fe elements in the dust clouds which is released from the solid state into the gas state under UV photons (Jiang et al. 2019).
They are the observational evidence of the dust sublimation in the model.

The dust in the nuclear region can not only produce a MIR flare by echo, but can also obscure the UV and optical radiation of a transient event if the dust is located on the line of sight.
Hence if the dust obscuration is severe, the dust can hide the transient event from us in optical surveys.
Hidden transient events in active galaxies maybe common, considering that more than half of active galaxies belong to type II or type 1.9 as their nuclei are severely obscured by dust.
The hidden transient events disturb the statistics of the transient events.
For an accurate statistics, methods other than optical surveys are required to detect the hidden transient events.
MIR radiation is little affected by dust extinction, and thus the hidden transient events in active galaxies can be found using accompanied MIR flares as probes.
An example is an IR flare named Arp 299-B AT1 (Mattila et al. 2018), which was discovered with ground-based near-infrared (NIR) observations and Spitzer MIR observations.
It occurred at the nucleus of Arp 299-B1, which hosts a Compton-thick AGN.
It caused the IR luminosity of the nucleus to rise continuously and reach the peak during $\sim$5 years, and then decrease at a lower rate.
It released a total IR energy of $>10^{52}$ erg over a duration of more than ten years.
The IR flare was accompanied by a weak optical variation discovered by HST that could be related to it.
The optical variation was so weak that it might have been missed if the IR flare had not been discovered first or the galaxy had not been observed by HST many times.
The IR flare was also accompanied by a transient radio jet, and based on which Mattila et al. demonstrated that the IR flare was caused by a TDE.

The WISE (Wright et al. 2010) survey is a powerful tool for finding the hidden transient events via accompanied MIR flares.
The WISE telescope has been conducting a repetitive all-sky survey since 2010.
The survey was named WISE, NEOWISE (Mainzer et al. 2011) and NEOWISE-R (Mainzer et al. 2014) at different time.
Every half a year, except for a hiatus between 2011 and 2013, WISE took multiple exposures during several days for a certain sky region.
The observations were initially carried out with four MIR filters with central wavelengths of 3.4, 4.6, 12 and 22 microns, named W1 to W4, and then with only W1 and W2 filters after the cryogen was exhausted in September 2010.
For all galaxies, WISE observations provide information of variabilities on time scales of hours, days and years.
There have been works that search for MIR flares with variability time scale of years using WISE data.
Wang et al. (2018) found 14 normal galaxies with long-term declining MIR radiations, and argued that the MIR variations may be caused by TDEs.
Assef et al. (2018) found 9 galaxies whose MIR LCs resemble that of W0948+0318, none of which showed optically detected transient event.
Yang et al. (2019) reported a MIR Flare in a Type II AGN J1657+2345 with no significant changes in the optical.

We are conducting a comprehensive search of MIR flares in nearby SDSS spectroscopic galaxies using WISE data (Jiang et al. in preparation).
We found several tens of MIR flares.
Among them some are in active galaxies, and a large fraction of them have no accompanied optical transient events detected.
We aim to extract as much information as possible about the transient events that caused the MIR flares based on only MIR data.
The information will help us understand the nature of the MIR flares.
We made an attempt using a MIR flare in MCG-02-04-026, which is a $z=0.03485$ edge-on disk galaxy located at $\alpha_{\rm J2000}=$01:20:47.99, $\delta_{\rm J2000}=$-08:29:18.4.
In this paper, we will present observations and data of the galaxy in Section 2, analyze the data in Section 3, model the MIR data and extract information about the primary transient event in Section 4, and finally explore the nature of the MIR flare in Section 5.
Throughout this paper, we use cosmological parameters of $H_0=67$ km s$^{-1}$ Mpc$^{-1}$, $\Omega_m=0.32$ and $\Omega_\Gamma=0.68$, resulting in a luminosity distance of 160 Mpc for MCG-02-04-026.

\section{Observations and Data}

In Table 1, we list the basic information of all the observations of which the data were used in this paper.
The information includes the time, the band, the instrument and the mode (photometric/spectroscopic).
We described the details of the observations and the data in the following.

\subsection{WISE photometric data}

As we have mentioned, WISE took multiple exposures of each target within several days every half a year.
We referred to the multiple exposures as an observation.
For MCG-02-04-026, data from 12 observations had been published.
We list the information of the 12 observations in Table 2.
The first two observations are in 2010, and the remaining are during 2013 and 2018.
The first observation took images with all four filters, and the remaining observations only took images with W1 and W2 filters.
In each band of each observation, there were 10--25 exposures and images of the same number were obtained.

The WISE pipeline has performed photometries of each target by fitting the images with point spread functions (PSF).
We collected the PSF magnitudes of MCG-02-04-026 from the NASA/IPAC InfRared Science Archive (IRSA)\footnote{https://irsa.ipac.caltech.edu/applications/Gator/}.
We removed the bad magnitudes with poor image quality (qi\_fact$=$0), charged particle hits (saa\_sep$<$0), scattered moon light (moon\_masked$=$1) or artifacts (cc\_flags$>$0), and there were 8-¨C12 good magnitudes left in each band of each observation.
We did not detect any variations in any band of any observation.
So we binned the magnitudes in each band of each observation by taken the mean value and the standard deviation of the magnitudes as the value and the error, respectively.
The binned magnitudes were listed in Table 2.

Meisner et al. (2018) provided time-resolved WISE/NEOWISE coadded images of each observation.
For MCG-02-04-026, the coadded images from the first ten observations had been published\footnote{https://portal.nersc.gov/project/cosmo/temp/ameisner/neo4/}.

\subsection{Time-domain photometric/imaging data and X-ray data}

We collected optical photometric data of MCG-02-04-026 from three transient surveys, Catalina Sky Survey (CSS, Drake et al. 2009), Palomar Transient Factory (PTF, Law et al. 2009), and All-Sky Automated Survey for Supernovae (ASAS-SN, Shappee et al. 2014; Kochanek et al. 2017)\footnote{We collected CSS photometries from CSS data releases 2 (http://nesssi.cacr.caltech.edu/DataRelease/), PTF photometries from IRSA, and ASAS-SN photometries from their database (https://asas-sn.osu.edu/).}.
MCG-02-04-026 was observed by CSS with $V_{\rm CSS}$ filter over 42 months (66 nights) during August 2005 and October 2013, by PTF with $R$ filter over 8 months (17 nights) during July 2009 and November 2012, and by ASAS-SN with $V$ filter over 52 months (409 nights) during October 2012 and November 2018.
For each survey, we binned the photometric data taken in each month to obtained a LC using the same method as we did for the WISE/NEOWISE data.
The binned LCs are listed in Table 3.

We collected UV imaging data and X-ray data of MCG-02-04-026.
We found archival UV imaging data from SDSS observation with SDSS-u filter in October 2009, GALEX observation with FUV and NUV filters in October 2013, and SWIFT/UVOT observation with UVW2 filter in May 2015.
We also found archival X-ray data from ROSAT/PSPC All-Sky Survey (RASS) observation taken between July 1990 and January 1991, and SWIFT/XRT observation taken in May 2015.
We applied for a new SWIFT target of opportunity observation of MCG-02-04-026 and the observation was taken in June 2018.
An image was obtained by SWIFT/UVOT with U filter and X-ray data was obtained by SWIFT/XRT.
We listed the basic information of the UV and X-ray observations in Table 4\footnote{We collected SDSS image from SDSS DR12 (https://www.sdss.org/dr12/), GALEX image from GALEX DR6 (http://galex.stsci.edu/GR6/), SWIFT/UVOT images, SWIFT/XRT event files, and ROSAT/PSPC event files from The High Energy Astrophysics Science Archive Research Center (HEASARC) (https://heasarc.gsfc.nasa.gov/)}.

\subsection{Optical and NIR Spectral Data}

We collected an archival optical spectrum of MCG-02-04-026 from SDSS-III's Baryon Oscillation Spectroscopic Survey (BOSS)\footnote{We downloaded the spectrum from SDSS DR12.}.
The spectrum was taken on November 2013 with a total exposure time of 7200 seconds, a wavelength coverage of 3600--10400 \AA\ and a spectral resolution of 1500--2500.

We took follow-up quasi-instantaneous optical and NIR spectroscopic observations of MCG-02-04-026 in January 2018.
The optical spectrum was taken on 2018-01-05 with the Double Spectrograph (DBSP) on the Hale 200-inch telescope at Palomar Observatory, and the NIR spectrum was taken on 2018-01-03 with the TripleSpec (TPSP) spectrograph on the same telescope.
For the DBSP observation, we used a D55 dichroic, a 600/4000 grating for the blue side, and a 316/7500 grating for the red side.
We adjusted the grating angles to obtain a nearly continuous wavelength coverage from 3200 to 10000 \AA\ except for a small gap of 5470--5490 \AA.
We used a slit with $1.0\arcsec$ width because the seeing FWHM was $\sim1.2\arcsec$, and oriented the slit in the north-south direction, which is close to the direction of the galaxy disk.
We took two exposures in order to remove cosmic rays.
The total exposure time was 20 minutes.
We took Fe-Ar and He-Ne-Ar lamp spectra for the wavelength calibration for blue and red sides, respectively.
The standard star for flux calibration was observed about 1 hour before the observation of MCG-02-04-026.
We reduced the data following a standard routine, and extracted a spectrum in 3$\arcsec$ aperture.
For the TPSP observation, we used a slit with 1.0$\arcsec$ width, which resulted in a spectral resolution of 2000--2500.
We used A-B-B-A dithering mode to better remove the sky emission lines.
The total exposure time was 40 minutes.
We reduced the data and extracted a spectrum with the specX package.

\subsection{Multi-band Photometry for Spectral Energy Distribution}

In order to generate the spectral energy distribution (SED) of MCG-02-04-026 before the MIR flare, we collected data observed by GALEX in 2003, by SDSS in 2009, by 2MASS in 1998, by WISE in 2010 (the first observation), by IRAS in 1983, and by VLA/FIRST in 1997.
We list the basic information of the observations in Table 5, including the facilities, the filters and their central wavelengths, and the dates.
We aimed to get the integrated flux of the entire galaxy to avoid the influence of different spatial resolutions of images in different bands.
For SDSS, 2MASS, IRAS and VLA/FIRST observations, we collected the integrated fluxes of MCG-02-04-026 or corresponding magnitudes from public catalogs, which are described in detail in Table 5.
And for GALEX and WISE observations, we measured the Petrosian aperture fluxes of the galaxy with Sextractor (v2.8.6, Bertin \& Arnouts. 1996).
We also list the magnitudes/flux in Table 5.

\section{Data Analysis}

\subsection{The MIR Flare}

We generated the MIR LCs of MCG-02-04-026 over the 12 WISE observations using the PSF magnitudes in W1 and W2 bands.
We display the LCs in Figure 1(a).
Before 2014, MCG-02-04-026's MIR radiation remained steady (W1$\sim$11.00 and W2$\sim$10.46).
A rise of the radiation was captured by the 7 observations during 2014 and 2017, indicating that there may be a flare in MCG-02-04-026.
The radiation rose rapidly between January and July in 2014, and kept rising to the peak (W1$=10.56\pm0.02$ and W2$=9.89\pm0.01$) in December 2015.
The radiation then dropped, and after 2017, it returned to the initial level with small variation.

We further confirmed the flare using image subtraction technic.
We did this using the WISE time-resolved coadded images from Meisner et al. (2018).
In each of W1 and W2 bands, we created a pre-flare image by combining the images from the first 3 observations, and a flare image by combining the images from 7 observations during July 2014 and July 2017, and then subtracted the pre-flare image from the flare image to obtain a residual image.
We show these images in Figure 2.
In the residual images, clearly excess radiations can be seen at the position of the galaxy, and meanwhile nearby objects show no significant excess radiations.

We measured the position of the flare by fitting the spatial brightness profile of the excess radiation with a two-dimensional Gaussian model.
The results showed that the offsets between the position of the flare and the galaxy center are only 0.12 and 0.10 pixels\footnote{1 pixel$=2.75\arcsec=1.92$ kpc} in W1 and W2 bands, respectively.
The errors to these offsets are dominated by systematic errors, which are at sub-pixel levels typically.
So the position of the flare coincides with the galaxy center, and we can constrain their angular distance in 1 pixel, and their projected distance in $\sim$2 kpc in physical size.
Besides, the flare behaved like a point source because the FWHM of the excess radiation (6.8$\arcsec$ in W1 and 7.4$\arcsec$ in W2) is consistent with those of nearby stars (6.1--7.1$\arcsec$ in W1 and 6.5--7.5$\arcsec$ in W2).

We measured the fluxes of the flare in each band of the 7 observations between July 2014 and July 2017.
We subtracted the pre-flare image which we obtained earlier from the coadded images in each band of each observation to get residual images.
Then we performed aperture photometry to the residual images.
Except for the random error obtained in following the routine method of aperture photometry, there must also be a systematic error, which can be caused by small position offset, PSF variation, and some other factors.
In order to measure the systematic error, we made a test using several tens of nearby stars.
We performed aperture photometry at the positions of the stars on the residual images.
For each star, we calculated the root square mean (RMS) of its fluxes on the 7 residual images in each band.
We found that the RMS values of the stars are generally greater than the values predicted by random errors.
So additional systematic errors, which are predicted to be $\sim$3\% of the stellar fluxes, are required to explain the RMS values.
Thus for the flux of the flare, a systematic error of 3\% of the flux of the galaxy nucleus measured in the pre-flare image, was considered.
The results, including the fluxes, the random and systematic errors, are listed in Table 6.

We generated the LCs and spectra of the flare, and display them in Figure 3(a) and Figure 3(b).
The LC of the flare shows similar pattern with the LC of MCG-02-04-026.
We saw a change in the spectral shape of the flare (Figure 3(b)).
The change can be seen more clearly from the variation of the MIR color, which can be expressed as the ratio between the flare's fluxes in W1 and W2 bands ($f_\nu$(W1)/$f_\nu$(W2), smaller for redder), which is plotted in Figure 3(c).
The MIR color of the flare was generally turning red.

We estimated the IR luminosity ($L_{\rm IR}$) and the total IR energy ($E_{\rm IR}$) of the flare as following.
First, because only the data in the W1 and W2 bands (their passbands are 2.8--3.8 $\mu$m and 4.0--5.3 $\mu$m, respectively) are available, we calculated the IR luminosity in a range of 2.8--5.3 $\mu$m ($L_{2.8-5.3\mu m}$), and considered it as the lower limit of $L_{\rm IR}$.
We fit the spectra of the flare with power-law models, which are expressed as:
\begin{equation}
f_\nu(\lambda) = \left\{
    \begin{array}{lr}
      { C_1 \lambda^{C_2},\ \lambda_1 < \lambda < \lambda_2 } \\
      { 0,\ \lambda < \lambda_1\ {\rm or}\ \lambda > \lambda_2 }
    \end{array} \right. \\
\end{equation}
where $C_1$ and $C_2$ are free parameters, and $\lambda_1=2.8$ $\mu$m and $\lambda_2=5.3$ $\mu$m.
The $L_{\rm 2.8-5.3\mu m}$ inferred from the power-law models are shown in Figure 3(d).
We obtained an energy of $7.4\times10^{50}$ erg by integrating the $L_{2.8-5.3\mu m}$ curves over time.
This value sets the lower limit of $E_{\rm IR}$.
Then we estimate the $L_{\rm IR}$ using blackbody models, because the IR spectrum of a similar MIR flare, Arp 299-B AT1 (Mattila et al. 2018), conforms to blackbody curve in a wider wavelength range (1.2 to 8 $\mu$m).
The blackbody models are expressed as:
\begin{equation}
f_\nu = \frac{ \pi B_\nu(T_{\rm BB}) 4\pi r_{\rm BB}^2 }{ 4\pi D_L^2 }
\end{equation}
where $T_{\rm BB}$ (blackbody temperature) and $r_{\rm BB}$ (blackbody radius) are free parameters.
The blackbody temperature (Figure 3(e)) was roughly dropping, corresponding to the variation in the spectral shape, and was about 1000 K in most of the time.
The blackbody radius (Figure 3(f)) was roughly increasing and reached a maximum value of $\sim$100 light-days in 2016.
Using blackbody models, we calculated blackbody luminosities (Figure 3(g)), and integrated them up to obtain a total energy of $1.8\times10^{51}$ erg.
Thus the total IR energy $E_{\rm IR}$ exceeds $\sim7\times10^{50}$ erg, and is likely to be on the order of $10^{51}$ erg.

\subsection{No Detection of Variability in Optical and UV}

We show the optical LCs of MCG-02-04-026 from CSS, PTF and ASAS-SN in Figure 1(b).
The mean magnitudes of the three LCs are different ($V_{\rm CSS}=14.75$ for CSS, $R=14.85$ for PTF and $V=15.27$ for ASAS-SN), because the bands and apertures for photometry are different.
We saw no internal variability in any of the three LCs.
Thus no optical flare corresponding to the MIR flare was recorded.
The typical error levels, which are calculated from the RMS of the errors for each LC, are 0.03, 0.05 and 0.04 magnitudes for CSS, PTF and ASAS-SN, respectively.
Thus any possible optical flare either fell into the gap between observations, or had a variability amplitude lower than 0.1 magnitude.

We check possible UV variation by subtracting UV images taken before 2014 from those taken after 2014, the year when the MIR flare began.
We used two pairs of images, each of which contains a pre-2014 image and a post-2014 image, as listed in Table 4.
In the first pair, the post-2014 image was taken by SWIFT/UVOT in 2015 with UVW2 filter (central wavelength $\lambda_c=2140$ \AA), and the pre-2014 image was a fake image created using GALEX images taken in 2003 with NUV ($\lambda_c=2340$ \AA) and FUV ($\lambda_c=1550$ \AA) filters.
We created the fake image as follows: for each spatial pixel, we calculated the flux at 2140 \AA\ by interpolating the fluxes at 2340 \AA\ and at 1550 \AA\ by assuming a spectral shape of a power law.
In the second pair, the pre-2014 image was taken by SDSS in 2009 with u filter ($\lambda_c=3560$ \AA), and the post-2014 image was taken in 2018 with U filter ($\lambda_c=3490$ \AA).
For each pair, the subtraction was done as follows: we first aligned the pair of images, and then blurred the image with higher spatial resolution by convolving with a Gaussian to let it have the same spatial resolution with the other image, and then multiplied one of the pair by a constant factor, and finally did subtraction.
The pre-2014, post-2014 and the residual images in the two pair are shown in Figure 4.
Neither of the residual images shows obvious excess radiation.

To conclude, neither optical nor UV flare corresponding to the MIR flare was detected based on available data.

\subsection{The Optical and NIR Spectral properties and possible spectral variations}

We show the two optical spectra of MCG-02-04-026, including a spectrum taken with BOSS in November 2013 (hereafter Y2013) and a spectrum taken with DBSP in January 2018 (hereafter Y2018), in Figure 5(a).
In both of the two spectra, the continua are dominated by starlight and prominent emission lines can be seen.
To inspect possible spectral variation caused by the MIR flare, we calculated the ratio between the Y2018 and Y2013 spectra (Figure 5(b)).
The ratio shows that the continuum varied little.
A wavelength-independent variation with amplitude of $\sim$5\% is possibly due to different apertures used when taken the two spectra (Figure 5(c)).
The ratio also shows that the flux around H$\alpha$ increased from 2013 to 2018, and we will explain this later.

Since we were concerned with the emission lines, we modeled the starlight and subtracted them from the observed spectra to obtain emission line spectra.
We modeled the starlight using a linear combination of principal component templates of spectra of simple stellar populations (SSP).
We used the templates from Lu et al. (2006), which were built from the library of SSP spectra of Bruzual \& Charlot (2003).
We fit the starlight model to the two spectra in regions which are free of emission lines.
In the fitting, the combination of the templates was shifted with a redshift parameter $z$(star), and broadened by convolving with a Gaussian with a velocity dispersion parameter $\sigma$(star), and reddened by multiplying by the Milky Way's extinction curve of Fitzpatric (1999) with a reddening parameter $E_{B-V}$(star).
We plot the starlight models in Figure 5(a), and list the parameters in Table 7.

In Figure 6, we show the emission line spectra from the Y2013 and Y2018 observations in two wavelength ranges, one is around H$\beta$+[OIII] and the other is around H$\alpha$.
In both of the two spectra, we detected a H$\alpha$ broad emission lines (BEL), and meanwhile detected no H$\beta$ BEL.
For each spectrum, we simultaneously modeled the H$\alpha$ BEL and narrow emission lines (NELs), including H$\beta$, [O III] $\lambda\lambda$4959, 5007, [O I] $\lambda$6300, [N II] $\lambda\lambda$6548, 6583, H$\alpha$ and [S II] $\lambda\lambda$6716, 6731.
In the model, H$\alpha$ BEL and all the NELs are assumed to have single Gaussian profiles.
The centroids and widths in term of velocity of all the NELs are tied.
The flux ratios of [N II] and [O III] doublet lines are fixed to be 3.
We plotted the emission line model in Figure 6 and listed the parameters in Table 7.

We can categorize MCG-02-04-026 as a Seyfert 1.9 galaxy according to the detection of H$\alpha$ BEL and no detection of H$\beta$ BEL.
We can also categorize it as a Low Ionization Nuclear Emission Region (LINER) galaxy using the ratios of NELs, which are listed in Table 7, according to the diagnostic diagram of Kewley et al. (2006).
The categorizations based on the two optical spectra are consistent, and indicate that MCG-02-04-026 is an active galaxy.
Note that there is a possible non-AGN interpretation of the BELs that the BELs might be newly generated by an event that caused the MIR flare.
This was unlikely to happen, because the Y2013 observation was taken 1.5 month before the WISE observation in January 2014 during which no MIR excess radiation was seen, and because the Y2018 observation was taken when the MIR flare had faded.
Even if the BELs were not caused by AGN, the narrow line ratios like those of LINERs are evidence of MCG-02-04-026 being an active galaxy, because LINERs are usually related to AGN (Ho 1999), and the narrow line fluxes do not change in a short time.

The model shows that the flux of H$\alpha$ BEL on the Y2018 spectrum increases for $\sim$30\% relative to that on the Y2013 spectrum, and the profile of H$\alpha$ BEL also varies.
Such a variation is commonly seen in active galaxies, so we were not interested in this.
The variation is consistent with the fact we found earlier that the flux around H$\alpha$ increased from 2013 to 2018.

The NIR spectrum of MCG-02-04-026 (Figure 5(d)) is dominated by continuum.
The only emission line that we can identify with certainty is Pa$\beta$ BEL\footnote{Notice that the Pa$\alpha$ and Pa$\gamma$+He I $\lambda$10830 emission lines, which are commonly seen in Seyfert galaxy's spectrum, fall into regions strongly affected by telluric absorptions.}.
We modeled the local continuum around Pa$\beta$ using a two-order polynomial (red line in Figure 5(d)) and show the emission line spectrum after subtracting the continuum in the right panel of Figure 6.
We also modeled the Pa$\beta$ BEL using single Gaussian and list the parameters in Table 7.

Generally there is dust on the line of sight to the nucleus of Seyfert 1.9 galaxy, causing the reddening of the nuclear emission.
We estimated the dust reddening to the nucleus of MCG-02-04-026 by comparing the observed to the intrinsic flux ratios of the BELs.
The flux ratios of H$\alpha$/H$\beta$ and H$\alpha$/Pa$\beta$ were used.
Although H$\beta$ BEL is not detected, we can estimated an upper limit of its flux and hence a lower limit of H$\alpha$/H$\beta$.
The upper limit of H$\beta$ BEL flux is estimated as follows.
First we added a H$\beta$ BEL component of which the center and width in terms of velocity was tied to that of H$\alpha$ BEL, and fit the emission line spectrum again to obtain a random error.
Next we estimated the systematic error introduced when subtracting the continuum: we chose different wavelengths where no actual emission lines exist on the emission line spectrum, and supposed that there were hypothetical BELs at these wavelengths, and measured the fluxes they would have, and estimated the systematic error by interpreting the RMS of the fluxes.
Then we calculated a total error by combining the random error and the systematic error.
Finally we obtained a 3$\sigma$ upper limit of the flux of the H$\beta$ BEL.
We list the upper limits from the two optical spectra in Table 7.
We also list the observed flux ratios of H$\alpha$/H$\beta$ and H$\alpha$/Pa$\beta$ from Y2013 and Y2018 observations in Table 8.
We adopted Case B values as the intrinsic flux ratios: a H$\alpha$/H$\beta$ value of 2.72 (Gaskell 2017), and a H$\alpha$/Pa$\beta$ value of 19 (Storey \& Hummer 1995)\footnote{We assumed that the broad line region has an electron temperature $T_e=15000$ K and an electron density $n_e=10^9$ cm$^{-3}$.}.
We estimated the nuclear dust reddening using three extinction curves, including a Milky Way's one from Fitzpatric (1999), a LMC averaged one from Misselt et al. (1999), and a SMC's one from Pei (1992).
The resultant $E_{\rm B-V}$ values are listed in Table 8.
The $E_{\rm B-V}$ values inferred from the two flux ratios from the observations in 2018 are consistent, and finally we adopted $E_{\rm B-V}$ value from H$\alpha$/Pa$\beta$ flux ratio.
The $E_{\rm B-V}$ value from the observations in 2018 meets with the lower limit set by the observation in 2013.
Thus there is no evidence that the nuclear dust reddening varied between 2013 and 2018.

\subsection{The X-ray properties and variations}

We extracted X-ray images around MCG-02-04-026 using data from ROSAT 1991, SWIFT 2015 and SWIFT 2018 observations.
We extracted in a range of 0.1--2.4 keV for ROSAT data, and a range of 0.3--8 keV for SWIFT data.
The images are shown in Figure 7(a).
We labeled the position of MCG-02-04-026 with red circles with radii of 1 arcmin.
Note that there is another X-ray source SDSS J012047.53-082629.5, which is a Seyfert 1 galaxy at $z=0.230$, 3 arcmin north to MCG-02-04-026 (labeled with blue circles).
The X-ray radiations from the two galaxies are blended in ROSAT data due to the poor spatial resolution, leading to a large uncertainty of the flux of MCG-02-04-026 in 1991.
Thus we only analyzed the data from the two SWIFT observations taken in 2015 and 2018 hereafter.

The net count rate in 0.3--8 keV reduced from $(3.9\pm0.4)\times10^{-2}$ counts s$^{-1}$ in 2015 to $(1.6\pm0.3)\times10^{-2}$ counts s$^{-1}$ in 2018, indicating a significant (4.6 $\sigma$) variation (a factor of $2.4\pm0.5$) in X-ray radiation of the galaxy.
For a further check on the variation, we extracted spectra from the two observations.
We show the two spectra in Figure 7(b), which were binned so that the data had at least 5 counts per bin.
We fit the X-ray spectra with {\it XSPEC} (Version 12.9.0) with $C$-statistics adopted.
In the fitting we considered the absorption from neutral gas in the Milky Way with $N_H=3.8\times10^{20}$ cm$^{-2}$ (Kalberla et al. 2005).
Simple absorbed power-law models can fit both the two spectra.
The models have three parameters including a photon index ($\Gamma$), an absorbing Hydrogen column density ($N_H$), and an intrinsic 2--10 keV luminosity $L_{\rm 2-10keV}$.
We list the parameters in Table 9.
The $\Gamma$ and $N_H$ for the two spectra can not be distinguished statistically.
Because no change in $\Gamma$ or $N_H$ was seen, we interpreted the variation in the net count rate as the variation in the X-ray luminosity.
Thus we refit the two spectra by tying $\Gamma$ and $N_H$.
The intrinsic 2--10 keV luminosities $L_{\rm 2-10keV}$ are several $10^{42}$ erg s$^{-1}$ in both observations.
The luminosities follow the correlation between X-ray and H$\alpha$ luminosities for Seyfert galaxies (e.g., Panessa et al. 2006).
While the luminosities are higher than the typical values of star forming galaxies (e.g., Mineo et al. 2014) and ultra-luminous X-ray sources (e.g., Stobbart et al. 2006).
Thus the X-ray radiation in MCG-02-04-026 can be related to the AGN.
However, there is also a possibility that the X-ray radiation could be generated by the event that caused the MIR flare, because both the two observations were taken after the MIR flare began.
Beside, the Hydrogen column density $N_H=(2.2\pm0.6)\times10^{22}$ cm$^{-2}$ falls between the typical values of Seyfert 1 and Seyfert 2 galaxies, and is consistent with those of Seyfert 1.9 galaxies (e.g., Koss et al. 2017).

\subsection{Pre-flare Spectral Energy Distribution}

In section 2.4, we generated a pre-flare SED of MCG-02-04-026 using photometric data taken before 2014.
We show the SED in UV to FIR bands in Figure 8(a).
The energy radiated by the galaxy is mainly in two wavelength ranges, one is the optical and NIR bands, and the other is the FIR band.
This is in line with disk galaxies with recent star formation.
We first fit the SED with a simple stellar radiation model.
We fit it with a python Code Investigating GALaxy Emission (CIGALE, Boquien et al. 2019), in which the radiation from stars and AGN, the dust attenuation, and dust radiation are considered simultaneously by assuming an energy balance.
We assumed that the stellar radiation consists of two components, one is starlight which is attenuated by dust, and the other is radiation of dust heated by stars.
We assumed a delayed star formation history (SF) expressed as equation (4) in Boquien et al. (2019) with three free parameters, including an age $t_0$, the time at which the SFR peaks $\tau$, and a normalization.
Using these parameters a stellar mass $M_{\rm star}$ and a star formation rate (SFR) can be inferred.
We used the SSP templates of Bruzual \& Charlot (2003), and assumed an initial mass function described in Salpeter (1955), and assumed a metallicity of 0.02.
We selected the dust attenuation law described in Charlot \& Fall (2000) with one free parameter which describes the V-band attenuation for the ISM $A_V$(ISM).
We used dust emission templates provided by Dale et al. (2014) with one free parameter $\alpha$.
The simple stellar radiation model is shown in Figure 8(a).
The model can fit the SED in optical, NIR and FIR bands.
However the model is lower than the observed SED in UV and MIR bands, indicating excess radiations in the SED relative to a simple stellar radiation model.

We then explored the nature of the excess radiation in the UV band.
We checked the GALEX images (Figure 4) and found that a large fraction of the UV flux comes from the outer region of the galaxy.
The UV radiation is more extended than the optical radiation.
The case is similar with those so-called ``extended UV disks'' (XUV-disk, e.g., Gil de Paz et al. 2005).
After visually inspecting the images, we found that the radiation of the XUV-disk contributes to almost all of the galaxy's flux in the FUV band, a large fraction of flux in the NUV band, a small fraction of flux in the u band, and negligible flux in bands with longer wavelengths.
So we decomposed the radiation of the galaxy by assuming that it comes from two components, a XUV-disk component and a Main-body component.
We decomposed it with a code TPHOT (Merlin et al. 2016).
We assumed that all the galaxy's flux in FUV band comes from the XUV-disk component, that all the flux in g band comes from the Main-body component, and that the fluxes in NUV and u bands come from both the two components.
We also assumed that the brightness profiles of each component in different bands are the same.
The results show that the XUV-disk component contributes 50\% and 14\% of the fluxes in NUV and u bands, respectively.
We fit the SED of the XUV-disk component (magenta pentagram in Figure 8(b)) in FUV to u bands with one other simple stellar radiation model.
The model is shown in cyan line in Figure 8(b) and the parameters are listed in Table 10.
The dust extinction of the XUV-disk component is little, and the corresponding dust radiation is negligible.
Using the model, we obtained the SFR in the XUV-disk component to be $0.5\pm0.1$ $M_\odot$ yr$^{-1}$.

We built an SED of the Main-body component in NUV to FIR bands.
The simple stellar radiation model cannot fit the SED as the $\chi^2$/d.o.f. is 33/10, and the excess radiation in the MIR bands still cannot be explained.
Excess radiation in the MIR band can be related to the radiation of dust heated by AGN.
Thus we added an AGN component in the model.
We chose AGN templates from Fritz et al. (2006), which are suitable for the radiation of obscured AGN.
We fixed some parameters as described in Table 10, leaving three free parameters of the component including the optical depth ($\tau$) at 9.7 $\mu$m, the angle ($\psi$) between the AGN axis and the line of sight, and a normalization.
An AGN luminosity can be inferred from the component.
After adding this AGN component, the model fit the data as $\chi^2$/d.o.f. is 6.6/7.
We show the model and its components in Figure 8(b), and list the parameters in Table 10.
We obtained the SFR in the Main-body component to be $2.8\pm0.3$ $M_\odot$ yr$^{-1}$, and the AGN luminosity ($L_{\rm AGN}$) to be $(3.5\pm0.6)\times10^{43}$ erg s$^{-1}$.
After combining the SFRs in the two components, we estimated the total SFR of MCG-02-04-026 to be $3.3\pm0.3$ $M_\odot$ yr$^{-1}$.

The VLA flux at 1.4 GHz is 3.6 mJy, corresponding to a radio power of $1.1\times10^{22}$ W Hz$^{-1}$.
Using the SED model from CIGALE, we calculated a 8--1000 $\mu$m luminosity of the galaxy to be $5\times10^{10}$ $L_\odot$.
The radio power and the 8--1000 $\mu$m luminosity of MCG-02-04-026 follow the radio-IR correlation from Bell (2003) for inactive galaxies, and hence the AGN's contribution to the radio radiation is unnecessary.

\subsection{The AGN nature}

There are three pieces of evidence of an AGN in MCG-02-04-026 before the MIR flare, including the H$\alpha$ BEL in the optical spectrum taken in 2013, the narrow line ratios which resemble those of LINERs, and the MIR excess in the pre-flare SED relative to the stellar radiation model.
After the MIR flare faded in 2017, both the BELs in the optical and NIR spectra taken in 2018 and the luminous ($L_{\rm 2-10keV}$ $>10^{42}$ erg s$^{-1}$) X-ray radiation in 2018 suggest that the nuclear activity was continuing, though we cannot rule out the possibility that these could be caused by the event which produced the MIR flare.
The radio radiation of MCG-02-04-026 is weak and can have a stellar origin, so the AGN is radio quiet.

The nuclear emission of MCG-02-04-026 is partially obscured because the BELs are reddened by dust and X-ray radiation is absorbed by gas.
We saw no evidence of variations of dust reddening and gas absorption.
Thus we considered them as invariable and adopted a dust reddening ($E_{\rm B-V}$) of 1.3--1.8, and hence an extinction ($A_V$) of 4.4--5.0, depending on the choice of extinction curves, and a Hydrogen column density ($N_H$) of $2.2\times10^{22}$ cm$^{-2}$.

We had obtained an AGN bolometric luminosity ($L_{\rm AGN}$) value of $10^{43.5}$ erg s$^{-1}$ after modelling the pre-flare SED in optical to IR band.
The $L_{\rm AGN}$ can also be estimated using X-ray luminosity $L_{\rm 2-10keV}$.
We chose the X-ray luminosity from the observation in 2018, because it was likely less affected by the event that caused the MIR flare, and adopted a bolometric correction of $\sim$20 from Hopkins et al. (2007).
We thus obtained one other $L_{\rm AGN}$ value of $10^{43.7}$ erg s$^{-1}$.
The two values are close, and hence we adopted the mean value of $L_{\rm AGN}\sim10^{43.6}$ erg s$^{-1}$ as the final value.

We estimated the mass of the central SMBH $M_{\rm BH}$ using the relation between the $M_{\rm BH}$ and the FWHM and luminosity of H$\alpha$ BEL (Greene \& Ho 2005).
The relation has an intrinsic scatter of 0.3 dex.
Because the AGN is partially obscured, a correction for dust extinction must be made when using the observed H$\alpha$ luminosity.
We estimated a correction factor to be 20--50, depending on the choice of extinction curves.
We calculated the $M_{\rm BH}$ using the two optical spectra separately.
The two values are consistent and we obtained a $M_{\rm BH}$ of $\sim10^{7.0\pm0.4} M_\odot$.
Note that the intrinsic scatter has been considered when calculating the error.

\section{The Nature of the MIR flare}

In this section we discuss the nature of the MIR flare in MCG-02-04-026.
We will interpret the observations of the flare with a dust echo model in Section 4.1, and discuss other possibilities in Section 4.2.

\subsection{Dust Echo of a Primary Transient Event}

Following some previous works on MIR flares in other galaxies (e.g., Jiang et al. 2017), we explore a possible scenario that the MIR flare in MCG-02-04-026 is the thermal reradiation from dust heated by UV-optical photons (i.e., dust echo) of a primary nuclear transient event.
Based on this scenario, we generated models of radiation of the dust echo by simulating the process, and during the simulation radiative transfer was involved.
We generated models in different situations with different parameters, and selected the optimal set of parameters so that the model fit the observations.

\subsubsection{A Brief Introduction to the Model}

Our model was based on the model of Lu et al. (2016, hereafter Lu16).
Briefly, Lu16's model considered a point source and the dust around the point source.
The point source was assumed to have a UV-optical luminosity $L(t)$ as a function of time $t$.
The dust was assumed to have a spherically-symmetric structure.
And the dust grains have the same initial radius $a_0$, and are uniformly distributed with number density $n_d$, spanning from inner radius $R_{\rm in}$ to outer radius $R_{\rm out}$.
When the UV-optical radiation of the transient event arrives at the inner surface of the dust structure, it heats, sublimates and destroys the dust grains there quickly.
After the radiation fully sublimates the innermost dust grains, it continues to sublimate the dust grains in the outer layers that have lost ``protection''.
The process goes on, and eventually stops at a distance of $R_{\rm sub}$ as the UV-optical radiation drops, and the outer dust grains can survive.
The UV-optical radiation is converted to IR reradiation by the dust grains it heats up.
The IR reradiation is attenuated by dust it passes through before it is finally received by the observer.
In addition, due to a so-called light travel time effect, the LC of the IR reradiation received by the observer is reshaped.
The effect causes a time delay of the IR reradiation relative to the UV-optical radiation because the light travel path of IR reradiation is longer than that of UV-optical radiation.
The effect also elongates the time duration of the IR reradiation because of difference in the lengths of the light travel paths of IR reradiations from dust grains at different positions.

There are three differences between our model and Lu16's model.
First, Lu16 assumed a FLAT form for the UV optical LC, i.e. a constant UV-optical radiation during the transient event, while we considered two more forms including a TDE form and a EXP form:
\begin{equation}
{\rm FLAT\ form:}\ \ \ L(t) =  \left\{
     \begin{array}{lr}
       { L_{\rm max},\ t_0<t<\tau } \\
       { 0,\ t<t_0 \vee t>\tau }
     \end{array} \right.
\end{equation}
\begin{equation}
{\rm TDE\ form:}\ \ \ L(t) =  \left\{
     \begin{array}{lr}
       { L_{\rm max} \left( 1+\frac{t-t_0}{\tau} \right)^{-5/3},\ t>t_0 } \\
       { 0,\ t<t_0 }
     \end{array} \right. \\
\end{equation}
\begin{equation}
{\rm EXP\ form:}\ \ \ L(t) =  \left\{
     \begin{array}{lr}
       { L_{\rm max}\ e^{-t/\tau},\ t>t_0 } \\
       { 0,\ t<t_0 }
     \end{array} \right.
\end{equation}
Second, we used approximate empirical formulas to calculate the variation of some physical quantities in the simulation.
These empirical formulas were obtained from results of rigorous simulations, and we will explain them in detail in the Appendix.
Using empirical formulas greatly reduces the computation time required to generate a model, while ensures validity of the results.
Third, we set $R_{\rm sub}$ a free parameter in the model.
The first reason is that the $R_{\rm sub}$ values obtained from the simulations may deviate from the reality, because there are too many parameters affecting $R_{\rm sub}$, especially because some of the parameters, such as dust chemical properties, are not fully considered in the model.
The second reason is that $R_{\rm sub}$ parameter has a great influence on the shape of the final MIR LCs.
In addition, the introduction of empirical formulas allowed us to set $R_{\rm sub}$ a free parameter.

\subsubsection{Fit to the data}

We fit the IR data of the MIR flare in MCG-02-04-026 with the simulations.
We calculated a total energy of the UV-optical radiation $E_{\rm tot}=\int_{t_0}^{\infty} L(t) {\rm d}t$, and set $t_0$, $L_{\rm max}$ and $E_{\rm tot}$ free parameters.
We assumed a 1:1 mixture of graphite and silicate, and used IR absorption coefficient data from Draine \& Lee (1984).
We tried four values of initial radius of grains ($a_0$) of 0.05, 0.10, 0.15 and 0.20 $\mu$m.
We fixed the value of $R_{\rm in}$ to be the sublimation radius corresponding to the AGN luminosity $L_{\rm AGN}\sim10^{43.6}$ erg s$^{-1}$ using the relation from Peterson (1997):
\begin{equation}
R_{\rm in} = 0.4\times \left( \frac{L_{\rm AGN}}{10^{45}\ \rm erg\ s^{-1}} \right)^{1/2}\ {\rm pc}
\end{equation}
Thus $R_{\rm in}$ is about 100 light-day.
We set $R_{\rm sub}$, $R_{\rm out}$ and $n_d$ free parameters.
Finally the model has 6 free parameters including $t_0$, $L_{\rm max}$, $E_{\rm tot}$, $R_{\rm sub}$, $R_{\rm out}$ and $n_d$, and a multi-selection parameter $a_0$ (4 selections).

For each form of UV-optical LC, we derived the parameters by minimizing $\chi^2$.
The overall minimum $\chi^2$ (10.8 for degree of freedom of 8) was obtained when the UV-optical LC has a TDE form, and the parameters are $a_0=0.15$ $\mu$m, $t_0{\rm (MJD)}=56740\pm30$, $E_{\rm tot}=9^{+4}_{-2}\times10^{51}$ erg, $L_{\rm max}=5^{+3}_{-2}\times10^{44}$ erg s$^{-1}$, $R_{\rm sub}=370\pm20$ light-day, $R_{\rm out}=600^{+400}_{-200}$ light-day and $n_d=6^{+20}_{-3}\times10^{-9}$ cm$^{-3}$.
We show the model yielding the overall minimum $\chi^2$ in Figure 9 by plotting the UV-optical LC of the primary transient event in Figure 9(a) and plotting the MIR LCs of the dust echo in Figure 9(b).
As can be seen in Figure 9(b), the model reproduces the observation of the MIR flare.

The model predicted that the primary transient event reached its peak luminosity around March 2014.
However, the optical V-band LC data from ASAS-SN observations taken around this time show no optical flares.
As the nuclear region of MCG-02-04-026 is severely obscured by dust with $A_V=4.4-5.0$, we checked that if the no detection of optical variation can be explained by dust obscuration.
We generated V-band LC models by assuming that the SED of the UV-optical radiation of the primary transient event is a blackbody curve with a temperature of $1.7\times10^4$ K, which is the temperature value of PS16dtm when its optical luminosity was around the peak level (Blanchard et al. 2017).
We generated two models, one with no obscuration, and the other with an obscuration of $A_V=4.7$.
We show the two models in Figure 9(c).
The models show that the transient event might cause a V-band magnitude variation of $>$1 if there were no obscuration, while it could only result in a variation of $\sim$0.04 under a severe dust obscuration.
We had obtained an error level of the V-band magnitude of MCG-02-04-026 from ASAS-SN to be $\sigma=0.04$ magnitude.
Thus the predicted V-band magnitude variation is below the 3$\sigma$ detection level.
Therefore the dust obscuration can explain the no detection of optical variation.
The dust obscuration can also explain the no detection of variation in the UV band where extinction is greater.

\subsubsection{Constraints to the Form of UV-optical LC}

We list the fitting results by using UV-optical LCs with each of the three forms in Table 11.
We found that the fits using TDE-form and EXP-form UV-optical LCs are much better than fits using FLAT-form UV-optical LC.
To show this clearly, we display the best-fitting models with the three forms of UV-optical LCs in Figure 10.
Although the model using UV-optical LCs with FLAT form can give MIR LCs with shapes roughly matching the observation, it cannot reproduce the observed variation in the MIR color at the same time, as can be seen in Figure 10(d).
Thus the data seems to support UV-optical LCs that decrease soon (no more than several months) after reaching the peak, rather than UV-optical LCs with plateaus whose time duration is as long as $\sim$1 year.

We should note that our dust echo model assumed a fast-evolving transient event during which the UV-optical radiation peaks in a short time scale.
Under this assumption, most of the energy of the UV-optical radiation is released in a short time, and the shape of MIR LC is roughly determined by the light travel time effect.
Another possible model is that the MIR flare could be caused by a slowly-evolving transient event whose UV-optical LC has a similar shape with the observed MIR LC.
We checked this possible model using a simulation.
The simulation was similar with the simulation described in section 4.1.1, except that we assumed a slowly-evolving transient event whose UV-optical LC is shown in red line in Figure 10(a), and except that we set the inner radius of the dust structure a constant.
The latter assumption is because that it is not necessary to consider the change in the inner radius of the dust due to sublimation when the peak luminosity of the transient event is close to the AGN luminosity, as argued by Almeyda et al. (2017)\footnote{In their Section 2.5}.
We show the MIR LCs and the variation in the MIR color of the dust echo radiation in the simulation in Figure 10(b) to 10(d).
We compared them with those of dust echo radiations produced by fast-evolving transient events with TDE-form and EXP-form UV-optical LCs.
The shapes of MIR LCs are similar, however the variations in the MIR color are completely different.
The dust echo radiation produced by a slowly-evolving transient event turns blue as it rises, and turns red as it drops.
While the dust echo radiation produced by a fast-evolving transient event keeps turning red both when it rises and drops.
The latter is consistent with the observation.
Therefore the MIR flare in MCG-02-04-26 was produced by a fast-evolving transient event rather than a slowly-evolving one.

\subsection{Other possibilities}

The MIR flare in MCG-02-04-026 is unlikely to be non-thermal radiation from a relativistic jet that is amplified by the beaming effect like those in blazers (e.g., Jiang et al. 2012), because the galaxy is radio quiet, and because we saw no dramatic variability on time scale of hours to days.
Thus it must originate from thermal radiation from dust, which was heated by energy released in a primary transient event.

The dust can be heated by radiation or by shock.
Here we discuss whether the dust can be heated by shock or not.
The shock may be produced by a wind or a jet that collides with gas surrounding the dust.
Here the wind refers to slow and scattered gas flow, and the jet refers to fast (close to the speed of light) and concentrated gas flow.
We first considered the case of the wind.
There are two facts.
One is that the black-body radius of the MIR flare radiation goes up to $\sim100$ light-days, indicating that the IR radiating dust is distributed in a vast space.
The other is that most of the IR energy was released in three to four years, indicating that the wind interacts with the IR radiating dust in different positions in a shorter time.
These two facts required that the velocity of the wind must be at least $\sim$0.1 c.
The ejecta of supernova or the outflows of TDE may have such a high velocity in their early stages, but will slow down to $\sim$0.01 c or lower velocities after expanding to larger space.
Thus it is unlikely that the dust is heated by shock produced by wind.
We then considered the case of the jet.
In this case, the IR energy comes from the kinetic energy of the jet.
From the observed IR luminosity, it can be estimated that the power of the jet exceeds $2\times10^{36}$ W.
Considering that the kinetic energy is not completely converted into thermal energy, the power of the jet needs to be higher.
Such a powerful jet would only been seen in radio galaxies (e.g., Merloni \& Heinz 2007).
MCG-02-04-026 is not a radio galaxy.
Thus it is unlikely that the dust is heated by shock produced by jet either.

Therefore, we concluded that the MIR flare is the dust echo of a primary transient event.

\section{The Origin of the Primary Transient Event}

Although the primary transient event was not seen directly due to dust obscuration, we can estimated its time scale and energy budget.
Most of the IR energy was released during $\sim$3 years.
Because the light travel time effect elongates the duration of dust echo radiation, the energy release time scale of the primary transient event that caused the MIR flare must be shorter.
Therefore, even the most conservative consideration can limit the energy release time scale to be less than $\sim3$ years.
We had obtained a total IR energy $E_{\rm IR}$ to be at least $\sim10^{51}$ erg.
Because not all of UV photons can be absorbed by dust and then reprocessed into IR photons, $E_{\rm IR}$ set a lower limit of the total energy of the primary transient event $E_{\rm tot}$.

On the nature of the primary transient event, we can rule out the possibility of normal SNe (that means not super-luminous supernova) because they are typically not bright enough in the MIR band.
For example, Szalai et al. (2019) collected MIR follow-up observations of hundreds of SNe, and found that among them the brightest SNe at 4.5$\mu$m is SN 2010jl with an absolute magnitude of $-22.8$, which is much dimmer than the MIR flare in MCG-02-04-026 with an absolute magnitude of $-25.2$ at 4.6$\mu$m.
We can also rule out the possibility of classical changing-look AGN events like those in Mrk 1018 and NGC 2627 because they evolve too slowly.
The time scale of the variation in the MIR radiation of these changing-look AGNs typically exceed $\sim$5 years (Sheng et al. 2017), much longer than that of the MIR flare in MCG-02-04-026.
We collected transient events meeting our constraints of energy budget and energy release time scale from the literatures.
These transient events had been listed in the introduction section.
Their interpretations include TDE, SLSNe, and enhancement of the existing accretion flow onto the SMBH.
Thus we considered these three kinds of events as the possible nature of the primary event causing the MIR flare in MCG-02-04-026.

We must point out that it is highly challenging to further distinguish the three possibilities.
The properties of the dust echo is not sensitive to the SED of primary event because sufficiently thick dust absorbs almost all UV-optical photons equally.
They are either not sensitive to the details of the UV-optical LC of the primary event, as the light travel time effect blurs the MIR LC.
Therefore, there are no other constraints on the nature of the primary event except for a rough energy budget and a rough energy release time scale based on only the MIR data.
We detected a significant X-ray variation between the two observations in 2015 and 2018 with a factor of $2.4\pm0.5$, and the X-ray luminosities in 2--10 keV in both the two observations exceeded $10^{42}$ erg s$^{-1}$.
Although such a X-ray variation is easy to be explained as a result of an enhancement of the existing accretion flow onto the SMBH, we cannot accordingly rule out the possibilities of TDE or SLSNe.
There are two reasons for this.
One is that the X-ray variation is not necessarily related to the MIR flare considering that X-ray variations with a factor of only 2.4 are common among AGN while MIR flares like that in MCG-02-04-026 are rare.
The other is that both TDE and SLSNe can be accompanied with strong X-ray radiation (e.g., TDE Swift J164449.3+573451, Bloom et al. 2011; Burrows et al. 2011; SNe-IIn, Dwarkadas \& Gruszko 2012).
Therefore, the origin of the MIR flare cannot be clarified based on the available X-ray observations.

Although we cannot make a definitive conclusion about the nature of the transient event that caused the MIR flare in MCG-02-04-026, here we provided a way to explore the nature of similar events in the future.
In our case, the first MIR excess was recorded (July 2014) 3 to 4 months after the predicted beginning time of the transient event.
Hence if future MIR surveys can have higher cadences than that of WISE, similar MIR flares will be caught in their earlier stages.
Although the optical messages of the transient events will be blocked by the dust, the messages in the NIR band, where extinction is small (e.g., $A_K\sim0.1A_V$ for the Milky Way's extinction curve), can still be received.
Therefore, after discovering MIR flares with MIR surveys, follow-up NIR observations should be taken as soon as possible to investigate the nature of the transient events.

\section{Conclusions}

We discovered a MIR flare using WISE data in the nearby edge-on disk galaxy MCG-02-04-026 at $z=0.03485$.
Before the MIR flare, MCG-02-04-026 is a dusty star-forming galaxy with a SFR of $\sim$3 $M_\odot$ yr$^{-1}$, and hosts a partially obscured ($E_{\rm B-V}=$1.3--1.8, $N_{\rm H}\sim2\times10^{22}$ cm$^{-2}$) active SMBH with $M_{\rm BH}\sim10^7$ $M_\odot$.
The MIR flare began in the first half of 2014, peaked around the end of 2015, and faded in 2017.
Its position coincided with the galaxy nucleus with an accuracy of $\sim$1 kpc.
During the three to four years, the MIR flare was roughly turning red.
The MIR flare released a total energy in the range of 2.8--5.3 $\mu$m ($E_{2.8-5.3\mu m}$) of $7.4\times10^{50}$ erg.
The total IR energy ($E_{\rm IR}$) must be higher than this value, and a better estimation of the $E_{\rm IR}$ can be obtained to be $\sim2\times10^{51}$ erg s$^{-1}$ using blackbody models.
Meanwhile neither optical nor UV variation corresponding to the MIR flare was detected based on available data.
X-ray net count rate changed by a factor of $2.4\pm0.5$ between two observations taken around the MIR flare, indicating a variation in the X-ray luminosity of the galaxy, however we were not sure whether it is related to the MIR flare or not.

With a dust echo model in which radiative transfer is involved, we interpreted the MIR flare as the reradiation of dust heated by UV-optical radiation from a primary nuclear transient event.
The model reproduces the MIR data, and also explains the variation of the MIR color.
The no detection of optical or UV variation can be explained as the dust obscuration to the galaxy nucleus.
We ruled out the possibility of radiation from relativistic jet or dust heated by shock.
Therefore we concluded that the nature of the MIR flare is the dust echo of the primary nuclear transient event.

The total energy of the primary transient event must be at least $\sim10^{51}$ erg, and most of the energy must be released during less than $\sim$3 years.
Such a transient event could be a TDE, a SLSNe, or an enhancement of the existing accretion onto the SMBH.
It is unlikely to be a normal SNe or a classical changing-look AGN event with a time scale of tens of years.

MCG-02-04-026 provides a good example that a MIR blind survey can help find hidden nuclear transient events.
Our analysis shows that using only MIR data, it is possible to constrain some properties of the transient event, such as the energy budget and the energy release time scale.
Future analysis of a large number of MIR flares like that seen in MCG-02-04-026 can improve our understanding towards hidden nuclear transient events, especially in active galaxies.

\begin{acknowledgements}
This work is supported by the NSFC grant (11421303, 11473025, 116203021, 11833007, NSF11033007), National Basic Research Program of China (973 Program, 2013CB834905), the SOC program (CHINARE2012-02-03), and Fundamental Research Funds for the Central Universities (WK 2030220010).
L.-M. Dou is also supported by NSFC grant 11833007, U1731104, jointly supported by Chinese Academy of Science and NSFC.
This work made use of data supplied by the UK Swift Science Data Centre at the University of Leicester.
This research also uses data obtained from the MAST.
This research uses data obtained through the Telescope Access Program (TAP), which has been funded by the Strategic Priority Research Program, the Emergence of Cosmological Structures (Grant No. XDB09000000), the National Astronomical Observatories, the Chinese Academy of Sciences, and the Special Fund for Astronomy from the Ministry of Finance.
\end{acknowledgements}

\clearpage

\begin{table}[!t]
\footnotesize
\caption{A summary of the observations of MCG-02-04-026.}
\begin{threeparttable}
\begin{tabular}{|c|c|c|c|c|c|c|c|c|c|}
\hline
\multirow{2}{1.0cm}{Year}  & Radio & FIR  & MIR  & \multicolumn{2}{c|}{NIR} & \multicolumn{2}{c|}{Optical} & UV & X-ray \\
\cline{2-10}
      & ---   & phot & phot & phot & spec & phot & spec & phot & ---   \\
\hline
1983  &       & IRAS &&&&&&&\\
\cline{1-1} \cline{3-3} \cline{10-10}
1991  &&&&&&&&& ROSAT \\
\cline{1-1} \cline{2-2} \cline{10-10}
1997  & VLA &&&&&&&&\\
\cline{1-1} \cline{2-2} \cline{5-5}
1998  &&&& 2MASS &&&&&\\
\cline{1-1} \cline{5-5} \cline{9-9}
2003  &&&&&&&& GALEX &\\
\cline{1-1} \cline{7-7} \cline{9-9}
2005--2008 &&&&&& CSS &&&\\
\cline{1-1} \cline{9-9}
2009  &&&&&& (2005-2013) && SDSS &\\
\cline{1-1} \cline{4-4} \cline{9-9}
2010  &&& WISE &&& SDSS &&&\\
\cline{1-1} \cline{4-4}
2011--2012 &&&&&& (2009) &&&\\
\cline{1-1} \cline{8-8}
2013  &&&&&& PTF & BOSS &&\\
\cline{1-1} \cline{4-4} \cline{8-8}
\specialrule{0em}{0.15pt}{0.15pt}
\rowcolor{mypink} 2014 &&&&&& (2009-2012) &&&\\
\cline{1-1} \cline{9-9} \cline{10-10}
\specialrule{0em}{0.15pt}{0.15pt}
\rowcolor{mypink} 2015 &&& Neo- &&& ASAS-SN && SWIFT/UVOT & SWIFT/XRT \\
\cline{1-1} \cline{9-9} \cline{10-10}
\specialrule{0em}{0.15pt}{0.15pt}
\rowcolor{mypink} 2016--2017 &&& WISE &&& (2012-2018) &&&\\
\cline{1-1} \cline{6-6} \cline{8-8} \cline{9-9} \cline{10-10}
\specialrule{0em}{0.15pt}{0.15pt}
2018 &&&&& TPSP & & DBSP & SWIFT/UVOT & SWIFT/XRT \\
\hline
\end{tabular}
\begin{tablenotes}
   \item ``phot'' stands for photometric observations, and ``spec'' stands for spectroscopic observations.
   We marked the observations during when the MIR flare began, evolved, and faded with pink background.
\end{tablenotes}
\end{threeparttable}
\label{tab1}
\end{table}

\begin{table}[!t]
\footnotesize
\caption{Logs of WISE observations of MCG-02-04-026 and binned PSF magnitudes.}
\begin{threeparttable}
\begin{tabular}{ccccccccc}
\hline
\hline
date & MJD & $N_{\rm image}$\tnote{a} & $N_{\rm good}$\tnote{a} & coadd\tnote{b} & \multicolumn{4}{c}{Magnitude} \\
 & & & & & W1 & W2 & W3 & W4 \\
\hline
2010-07-08 & 55386 & 13 & 12 & yes & $11.02\pm0.05$ & $10.48\pm0.04$ & $7.32\pm0.03$ &$5.38\pm0.07$\\
2010-12-31 & 55562 & 10 & 9  & yes & $10.99\pm0.05$ & $10.45\pm0.04$ & &\\
2014-01-02 & 56660 & 12 & 8  & yes & $10.99\pm0.02$ & $10.47\pm0.01$ & &\\
2014-07-09 & 56848 & 12 & 10 & yes & $10.71\pm0.02$ & $10.18\pm0.03$ & &\\
2015-01-04 & 57027 & 11 & 8  & yes & $10.68\pm0.02$ & $10.03\pm0.03$ & &\\
2015-07-06 & 57210 & 25 & 11 & yes & $10.62\pm0.02$ & $9.94\pm0.02$ & &\\
2015-12-29 & 57386 & 13 & 10 & yes & $10.58\pm0.05$ & $9.90\pm0.04$ & &\\
2016-07-07 & 57577 & 11 & 10 & yes & $10.65\pm0.03$ & $9.92\pm0.02$ & &\\
2016-12-22 & 57745 & 12 & 8  & yes & $10.89\pm0.05$ & $10.10\pm0.03$ & &\\
2017-07-06 & 57941 & 13 & 8  & yes & $10.97\pm0.03$ & $10.35\pm0.01$ & &\\
2017-12-18 & 58106 & 14 & 8  & no  & $11.09\pm0.03$ & $10.54\pm0.04$ & &\\
2018-07-08 & 58308 & 22 & 10 & no  & $10.90\pm0.04$ & $10.29\pm0.05$ & &\\
\hline
\end{tabular}
\begin{tablenotes}
   \item [a] Number of images taken during this observation, and number of good photometric data.
   \item [b] Whether a coadded image from this observation is available from Meisner et al. (2018) or not.
\end{tablenotes}
\end{threeparttable}
\label{tab2}
\end{table}

\begin{table}[!t]
\scriptsize
\caption{Logs of optical photometry of MCG-02-04-026 and binned LC.}
\begin{threeparttable}
\begin{tabular}{cc|cc|cc|cc}
\hline
\hline
Time (MJD) & Magnitude & Time (MJD) & Magnitude & Time (MJD) & Magnitude & Time (MJD) & Magnitude \\
\hline
\multicolumn{2}{c|}{CSS} & 2010-01(55216) & $14.85\pm0.04$ & 2012-11(56244) & $15.22\pm0.02$ & 2016-05(57531) & $15.33\pm0.06$ \\
\cline{1-2}
2005-08(53595) & $14.75\pm0.04$ & 2010-10(55489) & $14.72\pm0.02$ & 2012-12(56272) & $15.24\pm0.04$ & 2016-06(57554) & $15.18\pm0.03$ \\
2005-09(53627) & $14.74\pm0.02$ & 2010-11(55508) & $14.72\pm0.03$ & 2013-01(56317) & $15.18\pm0.06$ & 2016-07(57588) & $15.29\pm0.04$ \\
2005-10(53655) & $14.74\pm0.02$ & 2010-12(55536) & $14.72\pm0.03$ & 2013-06(56465) & $15.19\pm0.03$ & 2016-08(57614) & $15.23\pm0.04$ \\
2005-11(53691) & $14.74\pm0.02$ & 2011-01(55575) & $14.73\pm0.03$ & 2013-07(56488) & $15.20\pm0.02$ & 2016-09(57648) & $15.32\pm0.04$ \\
2005-12(53721) & $14.73\pm0.02$ & 2011-07(55774) & $14.82\pm0.04$ & 2013-08(56520) & $15.21\pm0.04$ & 2016-10(57677) & $15.25\pm0.02$ \\
2006-08(53952) & $14.82\pm0.04$ & 2011-09(55827) & $14.77\pm0.03$ & 2013-09(56553) & $15.30\pm0.06$ & 2016-11(57709) & $15.27\pm0.02$ \\
2006-09(54003) & $14.72\pm0.03$ & 2011-10(55857) & $14.73\pm0.02$ & 2013-10(56571) & $15.22\pm0.02$ & 2016-12(57742) & $15.30\pm0.03$ \\
2006-10(54025) & $14.73\pm0.02$ & 2011-11(55887) & $14.75\pm0.02$ & 2013-12(56641) & $15.25\pm0.02$ & 2017-01(57766) & $15.30\pm0.03$ \\
2006-11(54059) & $14.73\pm0.02$ & 2011-12(55923) & $14.74\pm0.03$ & 2014-01(56676) & $15.28\pm0.05$ & 2017-05(57901) & $15.11\pm0.06$ \\
2006-12(54088) & $14.71\pm0.02$ & 2012-10(56215) & $14.74\pm0.02$ & 2014-06(56831) & $15.28\pm0.05$ & 2017-06(57922) & $15.26\pm0.06$ \\
2007-06(54266) & $14.84\pm0.04$ & 2012-11(56233) & $14.76\pm0.03$ & 2014-07(56852) & $15.29\pm0.05$ & 2017-07(57952) & $15.27\pm0.03$ \\
2007-08(54333) & $14.82\pm0.04$ & 2013-01(56305) & $14.69\pm0.03$ & 2014-08(56888) & $15.28\pm0.02$ & 2017-08(57980) & $15.31\pm0.03$ \\
2007-09(54352) & $14.72\pm0.02$ & 2013-07(56478) & $14.81\pm0.03$ & 2014-09(56916) & $15.29\pm0.04$ & 2017-09(58014) & $15.28\pm0.03$ \\
2007-10(54386) & $14.72\pm0.02$ & 2013-09(56557) & $14.74\pm0.02$ & 2014-10(56946) & $15.29\pm0.02$ & 2017-10(58047) & $15.23\pm0.03$ \\
2007-11(54408) & $14.70\pm0.03$ & 2013-10(56585) & $14.73\pm0.02$ & 2014-11(56977) & $15.24\pm0.02$ & 2017-11(58071) & $15.26\pm0.04$ \\
\cline{3-4}
2008-01(54468) & $14.72\pm0.03$ & \multicolumn{2}{c|}{PTF} & 2014-12(57006) & $15.27\pm0.04$ & 2017-12(58099) & $15.23\pm0.05$ \\
\cline{3-4}
2008-08(54684) & $14.81\pm0.04$ & 2009-07(55044) & $14.75\pm0.04$ & 2015-01(57035) & $15.25\pm0.02$ & 2018-01(58130) & $15.31\pm0.04$ \\
2008-09(54722) & $14.73\pm0.02$ & 2009-08(55061) & $14.74\pm0.02$ & 2015-05(57164) & $15.31\pm0.04$ & 2018-06(58288) & $15.32\pm0.03$ \\
2008-10(54765) & $14.77\pm0.03$ & 2009-09(55088) & $14.74\pm0.02$ & 2015-06(57188) & $15.26\pm0.04$ & 2018-07(58319) & $15.35\pm0.02$ \\
2008-11(54787) & $14.74\pm0.02$ & 2009-11(55152) & $14.74\pm0.02$ & 2015-07(57222) & $15.26\pm0.04$ & 2018-08(58349) & $15.31\pm0.02$ \\
2009-01(54833) & $14.82\pm0.03$ & 2010-09(55460) & $14.73\pm0.02$ & 2015-08(57253) & $15.30\pm0.04$ & 2018-09(58371) & $15.28\pm0.01$ \\
2009-08(55060) & $14.69\pm0.03$ & 2011-12(55903) & $14.82\pm0.04$ & 2015-09(57280) & $15.26\pm0.04$ & 2018-10(58413) & $15.31\pm0.03$ \\
2009-09(55099) & $14.72\pm0.03$ & 2012-10(56228) & $14.72\pm0.03$ & 2015-10(57313) & $15.23\pm0.03$ & 2018-11(58443) & $15.40\pm0.03$ \\
2009-10(55127) & $14.73\pm0.02$ & 2012-11(56235) & $14.73\pm0.02$ & 2015-11(57339) & $15.28\pm0.03$ &  &  \\
\cline{3-4}
2009-11(55148) & $14.73\pm0.02$ & \multicolumn{2}{c|}{ASAS-SN} & 2015-12(57372) & $15.27\pm0.03$ &  &  \\
\cline{3-4}
2009-12(55182) & $14.74\pm0.03$ & 2012-10(56220) & $15.24\pm0.02$ & 2016-01(57401) & $15.23\pm0.04$ &  &  \\
\hline
\end{tabular}
\end{threeparttable}
\label{tab3}
\end{table}

\begin{table}[!t]
\footnotesize
\caption{Logs of UV imaging and X-ray observations of MCG-02-04-026.}
\begin{threeparttable}
\begin{tabular}{ccccccc}
\hline
\hline
Facility   & Band   & $\lambda$/E\tnote{a} & date  &  $T_{\rm exp}$ & Note \\
\hline
GALEX      & FUV    & 1550 \AA  & 2003-10-05 & -- & \multirow{2}{2.2cm}{Pair1, pre-2014}\\
GALEX      & NUV    & 2340 \AA  & 2003-10-05 & -- & \\
SDSS       & u      & 3560 \AA  & 2009-10-05 & -- & Pair2, pre-2014\\
SWIFT/UVOT & UVW2   & 2140 \AA  & 2015-05-23 & -- & Pair1, post-2014\\
SWIFT/UVOT & U      & 3490 \AA  & 2018-06-19 & -- & Pair2, post-2014\\
ROSAT/PSPC & X-ray  & 0.1--2.4 keV & 1990       & 0.93 ks & \\
SWIFT/XRT  & X-ray  & 0.3--8 keV   & 2015-05-23 & 2.4 ks  & \\
SWIFT/XRT  & X-ray  & 0.3--8 keV   & 2018-06-19 & 1.8 ks  & \\
\hline
\end{tabular}
\begin{tablenotes}
   \item [a] The central wavelengths for UV filters and the available wavelength ranges for X-ray observations.
\end{tablenotes}
\end{threeparttable}
\label{tab4}
\end{table}

\begin{table}[!t]
\footnotesize
  \caption{Photometric data of MCG-02-04-026 to construct a pre-flare SED.}
  \begin{threeparttable}
  \begin{tabular}{ccccccc}
  \hline
  \hline
  Facility/filter & $\lambda_{\rm c}$\tnote{a} & Mag/Flux & Date (UT) & Ref1\tnote{b} & Ref2\tnote{c} \\
  \hline
  GALEX/FUV & 0.155 $\mu$m & $18.85\pm0.08$ mag            & 2003-10-05 & 1 & -\\
  GALEX/NUV & 0.234 $\mu$m & $18.63\pm0.04$ mag            & 2003-10-05 & 1 & -\\
  SDSS/u    & 0.356 $\mu$m & $17.42\pm0.02$ mag\tnote{d}   & 2009-10-05 & 2 & 3\\
  SDSS/g    & 0.472 $\mu$m & $15.687\pm0.003$ mag\tnote{d} & 2009-10-05 & 2 & 3\\
  SDSS/r    & 0.619 $\mu$m & $14.807\pm0.003$ mag\tnote{d} & 2009-10-05 & 2 & 3\\
  SDSS/i    & 0.750 $\mu$m & $14.320\pm0.003$ mag\tnote{d} & 2009-10-05 & 2 & 3\\
  SDSS/z    & 0.896 $\mu$m & $13.901\pm0.004$ mag\tnote{d} & 2009-10-05 & 2 & 3\\
  2MASS/J   & 1.24 $\mu$m  & $12.61\pm0.03$ mag\tnote{e}   & 1998-10-30 & 4 & 5\\
  2MASS/H   & 1.66 $\mu$m  & $11.77\pm0.04$ mag\tnote{e}   & 1998-10-30 & 4 & 5\\
  2MASS/K   & 2.16 $\mu$m  & $11.31\pm0.04$ mag\tnote{e}   & 1998-10-30 & 4 & 5\\
  WISE/W1   & 3.35 $\mu$m  & $10.64\pm0.02$ mag   & 2010-07-09 & 6 & - \\
  WISE/W2   & 4.60 $\mu$m  & $10.21\pm0.02$ mag   & 2010-07-09 & 6 & - \\
  WISE/W3   & 11.6 $\mu$m  & $6.87\pm0.02$ mag    & 2010-07-09 & 6 & - \\
  WISE/W4   & 22.1 $\mu$m  & $5.20\pm0.03$ mag    & 2010-07-09 & 6 & - \\
  IRAS/60$\mu$m  & 61.5 $\mu$m & $0.51\pm0.05$ Jy\tnote{f} & 1983       & 7 & 8 \\
  IRAS/100$\mu$m & 102 $\mu$m & $1.27\pm0.13$ Jy\tnote{f}  & 1983       & 7 & 8 \\
  VLA/L     & 21 cm        & $3.68\pm0.16$ mJy\tnote{g}    & 1997-03-02 & 9 & 10 \\
  \hline
  \end{tabular}
  \begin{tablenotes}
    \item [a] The central wavelengths of the filters.
    \item [b] The references for the facilities: (1) Martin et al. (2005); (2) York et al. (2000); (4) Skrutskie et al. (2006); (6) Wright et al. (2010); (7) Neugebauer et al. (1984); (9) Becker et al. (1995).
    \item [c] The references for the magnitudes or fluxes: (3) Alam et al. (2015); (5) Jarrett et al. (2000); (8) Beichman et al. (1988); (10) Helfand et al. (2015). For GALEX and WISE, the magnitudes were measured from the images by us.
    \item [d] The magnitudes resulting from modelling the galaxy's brightness profile (modelMag), which were collected from the SDSS photometric catalog from SDSS Data Release 12.
    \item [e] The magnitudes measured in the ``total'' aperture defined in Jerrett et al. (2000), which were collected from the 2MASS extended source catalog.
    \item [f] The flux densities from the IRAS point source catalog.
    \item [g] The integrated flux density from the VLA/FIRST source catalog.
  \end{tablenotes}
  \end{threeparttable}
  \label{tab5}
\end{table}

\begin{table}[!t]
\scriptsize
\renewcommand\arraystretch{1.3}
\caption{The excess fluxes in MIR and the parameters of blackbody curves that match them.}
\begin{threeparttable}
\begin{tabular}{cccccccccc}
\hline
\hline
Date &MJD & Phase &$f_\nu$W1 &$f_\nu$W2 &$f_\nu$W1/$f_\nu$W2 &$T_{\rm BB}$  &$r_{\rm BB}$ &$L_{2.8-5.3\mu m}$ &$L_{\rm BB}$ \\
    &day & day   &mJy       &mJy       & & K & light-day & \multicolumn{2}{c}{$10^{42}$ erg s$^{-1}$} \\
\hline
2014-07-09 &56848 &0    &$3.71\pm0.13\pm0.36$ &$3.56\pm0.13\pm0.33$ &$1.03^{+0.17}_{-0.13}$ &$1400^{+370}_{-220}$ &$28^{+8}_{-8}$    &$5.7\pm0.5$  &$15^{+5}_{-3}$\\
2015-01-04 &57027 &173  &$4.92\pm0.14\pm0.36$ &$5.65\pm0.22\pm0.33$ &$0.87^{+0.11}_{-0.08}$ &$1140^{+170}_{-110}$ &$48^{+10}_{-9}$   &$8.0\pm0.5$  &$18.4^{+2.1}_{-1.3}$\\
2015-07-06 &57210 &350  &$5.32\pm0.13\pm0.36$ &$7.68\pm0.14\pm0.33$ &$0.69^{+0.06}_{-0.06}$ &$920^{+80}_{-60}$    &$79^{+12}_{-11}$  &$9.4\pm0.5$  &$21.4^{+0.9}_{-0.9}$\\
2015-12-29 &57386 &520  &$6.03\pm0.15\pm0.36$ &$7.71\pm0.17\pm0.33$ &$0.78^{+0.07}_{-0.06}$ &$1020^{+90}_{-70}$   &$66^{+10}_{-9}$   &$10.1\pm0.5$ &$22.8^{+1.2}_{-1.0}$\\
2016-07-07 &57577 &704  &$5.04\pm0.13\pm0.36$ &$8.04\pm0.14\pm0.33$ &$0.63^{+0.05}_{-0.05}$ &$850^{+60}_{-60}$    &$95^{+15}_{-12}$  &$9.3\pm0.5$  &$22.0^{+1.0}_{-0.9}$\\
2016-12-22 &57745 &867  &$2.04\pm0.13\pm0.36$ &$4.56\pm0.13\pm0.33$ &$0.45^{+0.09}_{-0.09}$ &$680^{+80}_{-80}$    &$120^{+50}_{-30}$ &$4.4\pm0.4$  &$13.4^{+2.5}_{-1.5}$\\
2017-07-06 &57941 &1056 &$1.11\pm0.13\pm0.36$ &$2.11\pm0.11\pm0.33$ &$0.54^{+0.23}_{-0.21}$ &$760^{+250}_{-190}$  &$60^{+70}_{-30}$  &$2.2\pm0.4$  &$6.3^{+2.1}_{-1.1}$\\
\hline
\end{tabular}
\end{threeparttable}
\label{tab6}
\end{table}

\begin{table}[!t]
\footnotesize
\caption{Parameters for modelling the optical and NIR spectra.}
\begin{threeparttable}
\begin{tabular}{ccc}
\hline
\hline
Parameter & 2013 & 2018 \\
\hline
\multicolumn{3}{c}{Starlight} \\
\hline
 $z$(star) & $0.03485\pm0.00002$ & $0.03485\pm0.00003$ \\
 $\sigma$(star)      & $90\pm5$ & $130\pm10$ \\
 $E_{\rm B-V}$(star) & $0.31\pm0.01$ & $0.30\pm0.01$ \\
\hline
\multicolumn{3}{c}{BEL}\\
\hline
 H$\alpha$ velocity & $460\pm10$  & $363\pm8$ \\
 H$\alpha$ FWHM     & $2500\pm20$ & $2230\pm15$ \\
 H$\alpha$ flux     & $2250\pm20$ & $2920\pm30$ \\
 Pa$\beta$ velocity & -           & $550\pm140$ \\
 Pa$\beta$ FWHM     & -           & $3100\pm300$ \\
 Pa$\beta$ flux     & -           & $1230\pm110$ \\
 H$\beta$ flux      & $<160$      & $<220$ \\
\hline
\multicolumn{3}{c}{NEL}\\
\hline
 velocity                   & $-50\pm1$ & $-51\pm1$ \\
 FWHM                       & $288\pm2$ & $326\pm3$ \\
 H$\beta$                   & $127\pm3$ & $108\pm7$ \\
 $\rm{[OIII]\ \lambda5007}$ & $348\pm5$ & $341\pm7$ \\
 $\rm{[OI]\ \lambda6300}$   & $66\pm3$  & $51\pm5$ \\
 H$\alpha$                  & $724\pm9$ & $687\pm11$ \\
 $\rm{[NII]\ \lambda6583}$  & $675\pm9$ & $661\pm10$ \\
 $\rm{[SII]\ \lambda6716}$  & $246\pm4$ & $227\pm5$ \\
 $\rm{[SII]\ \lambda6731}$  & $193\pm4$ & $185\pm5$ \\
\hline
\multicolumn{3}{c}{NEL ratio}\\
\hline
 $\rm{[OIII]/H\beta}$ & $2.74\pm0.07$   & $3.16\pm0.20$ \\
 $\rm{[NII]/H\alpha}$ & $0.93\pm0.02$   & $0.96\pm0.02$ \\
 $\rm{[SII]/H\alpha}$ & $0.61\pm0.01$   & $0.60\pm0.01$ \\
 $\rm{[OI]/H\alpha}$  & $0.091\pm0.004$ & $0.074\pm0.007$ \\
\hline
\end{tabular}
\begin{tablenotes}
    \item All the velocities, velocity dispersions and FWHMs are in unit of km s$^{-1}$.
    All the fluxes are in unit of $10^{-17}$ erg s$^{-1}$ cm$^{-2}$.
\end{tablenotes}
\end{threeparttable}
\label{tab7}
\end{table}

\begin{table}[!t]\footnotesize
\caption{The reddening to the BELs.}
\begin{threeparttable}
\begin{tabular}{c|c|cc}
\hline
\hline
\multicolumn{2}{c|}{Parameter} & 2013 & 2018 \\
\hline
\multirow{3}{1.2cm}{H$\alpha$/H$\beta$} & observed & $>$14.1 & $>$13.3 \\
   & intrinsic      & \multicolumn{2}{c}{2.72} \\
   & ratio          & $>$5.2 & $>$4.9 \\
\hline
\multirow{3}{1.2cm}{H$\alpha$/Pa$\beta$} & observed & - & $2.4\pm0.2$  \\
   & intrinsic & \multicolumn{2}{c}{19} \\
   & ratio     & -  & $0.12\pm0.01$ \\
\hline
\multirow{3}{1.2cm}{$E_{\rm B-V}$} & MW & $>$1.4 & $1.37\pm0.06$\tnote{a} \\
   & LMC average & $>$1.4 & $1.25\pm0.05$\tnote{a} \\
   & SMC         & $>$1.8 & $1.78\pm0.07$\tnote{a} \\
\hline
\end{tabular}
\begin{tablenotes}
    \item [a] The values converted from H$\alpha$/Pa$\beta$ flux ratios are adopted.
\end{tablenotes}
\end{threeparttable}
\label{tab8}
\end{table}

\begin{table}[!t]\footnotesize
\caption{Parameters for modelling the Swift spectra.}
\begin{threeparttable}
\begin{tabular}{c|cccc}
\hline
\hline
Parameter                                    & 2015 & 2018 & \multicolumn{2}{c}{2015\&2018 (tied)} \\
\hline
$\Gamma$                     & $2.8\pm0.6$ & $3.3\pm1.5$ & \multicolumn{2}{c}{$2.9\pm0.6$} \\
$N_H$ ($10^{22}$ cm$^{-2}$)  & $2.1\pm0.7$ & $2.5\pm1.6$ & \multicolumn{2}{c}{$2.2\pm0.6$} \\
$L_{\rm 2-10\ keV}$ ($10^{42}$ erg s$^{-1}$)  & $5.4^{+0.3}_{-2.4}$ & $<2.2$ & $5.2^{+0.3}_{-2.5}$ & $2.5^{+0.2}_{-1.2}$ \\
Cstat/d.o.f.                 & 7.7/14 & 2.8/2 & \multicolumn{2}{c}{10.7/18} \\
\hline
\end{tabular}
\end{threeparttable}
\label{tab9}
\end{table}

\begin{table}[!t]\footnotesize
\caption{Parameters for modelling the pre-flare SED with CIGALE.}
\begin{threeparttable}
\begin{tabular}{ccc}
\hline
\hline
\multicolumn{2}{c}{Parameter}   & Value \\
\hline
\multicolumn{3}{c}{Stellar radiation (XUV-disk)}\\
\hline
age             & free     & $40\pm30$ Myr \\
$\tau$          & free     & $60\pm30$ Myr\\
$M_{\rm star}$  & inferred & $(1.7\pm1.1)\times10^{7}$ $M_\odot$ \\
SFR             & inferred & $0.5\pm0.1$ $M_\odot$ yr$^{-1}$\\
$A_V$(ISM)      & free     & $<0.03$ mag\\
$\alpha$        & fixed    & 2 \\
\hline
\multicolumn{3}{c}{Stellar radiation (Mainbody)} \\
\hline
$t_0$           & free     & $5500\pm700$ Myr \\
$\tau$          & free     & $900\pm130$ Myr\\
$M_{\rm star}$  & inferred & $(1.3\pm0.1)\times10^{11}$ $M_\odot$ \\
SFR             & inferred & $2.8\pm0.3$ $M_\odot$ yr$^{-1}$\\
$A_V$(ISM)      & free     & $1.22\pm0.06$ mag\\
$\alpha$        & free     & $2.1\pm0.2$ \\
\hline
\multicolumn{3}{c}{AGN radiation}\\
\hline
$r$\tnote{a}      & fixed    & 60 \\
$\beta$\tnote{a}  & fixed    & $-0.5$ \\
$\gamma$\tnote{a} & fixed    & 4.0 \\
$\theta$\tnote{a} & fixed    & 100 \\
$\tau$          & free     & $1.1\pm0.6$ \\
$\psi$\tnote{b} & free     & $70\pm10$ deg \\
$L_{\rm AGN}$   & inferred & $(3.5\pm0.6)\times10^{43}$ erg s$^{-1}$\\
\hline
\end{tabular}
\begin{tablenotes}
  \item [a] The fixed parameters in the model are: the ratio of the maximum to minimum radii of the dust torus $r=60$, parameters describing the dust density distribution $\beta=-0.5$ and $\gamma=4$, and the opening angle of the dust torus $\theta=100^\circ$.
  \item [b] The $\psi$ value derived from the software is $20\pm10$ deg because of an error in Fritz et al. (2006).
\end{tablenotes}
\end{threeparttable}
\label{tab10}
\end{table}


\begin{table}[!t]
\footnotesize
\caption{Parameters for modelling the MIR data with the dust echo model.}
\begin{threeparttable}
 \begin{tabular}{c|c|c|c}
  \hline
  \hline
  \multirow{2}{1.2cm}{Parameters}   & \multicolumn{3}{c}{Form of $L(t)$} \\
  \cline{2-4}
                                             & TDE   & EXP   & FLAT \\
  \hline
  $a_0$ ($\mu$m)                             & 0.15            & 0.10             & 0.10 \\
  $T_0$ (MJD)                                & $56740\pm30$    & $56710\pm30$     & $56660\pm30$ \\
  log $L_{\rm max}$ ($10^{44}$ erg s$^{-1}$) & $5^{+3}_{-2}$   & $6^{+2}_{-1}$    & $2^{+2}_{-1}$ \\
  log $E_{\rm tot}$ ($10^{51}$ erg)          & $9^{+4}_{-2}$   & $12^{+5}_{-3}$   & $7^{+4}_{-2}$ \\
  $R_{\rm sub}$ (100 light-day)              & $3.7\pm0.2$     & $3.9\pm0.2$      & $4.3\pm0.2$ \\
  $R_{\rm out}$ (100 light-day)              & $6^{+4}_{-2}$   & $9^{+4}_{-2}$    & $8^{+4}_{-2}$ \\
  $n_d$ ($10^{-9}$ cm$^{-3}$)                & $6^{+20}_{-3}$  & $30^{+100}_{-10}$& $20^{+60}_{-10}$ \\
  \hline
  $\chi^2$/d.o.f.                            & 10.8/8 & 12.0/8 & 47.4/8 \\
  \hline
 \end{tabular}
\end{threeparttable}
\label{tab11}
\end{table}

\clearpage

\begin{figure}
\centering{
 \includegraphics[scale=0.98]{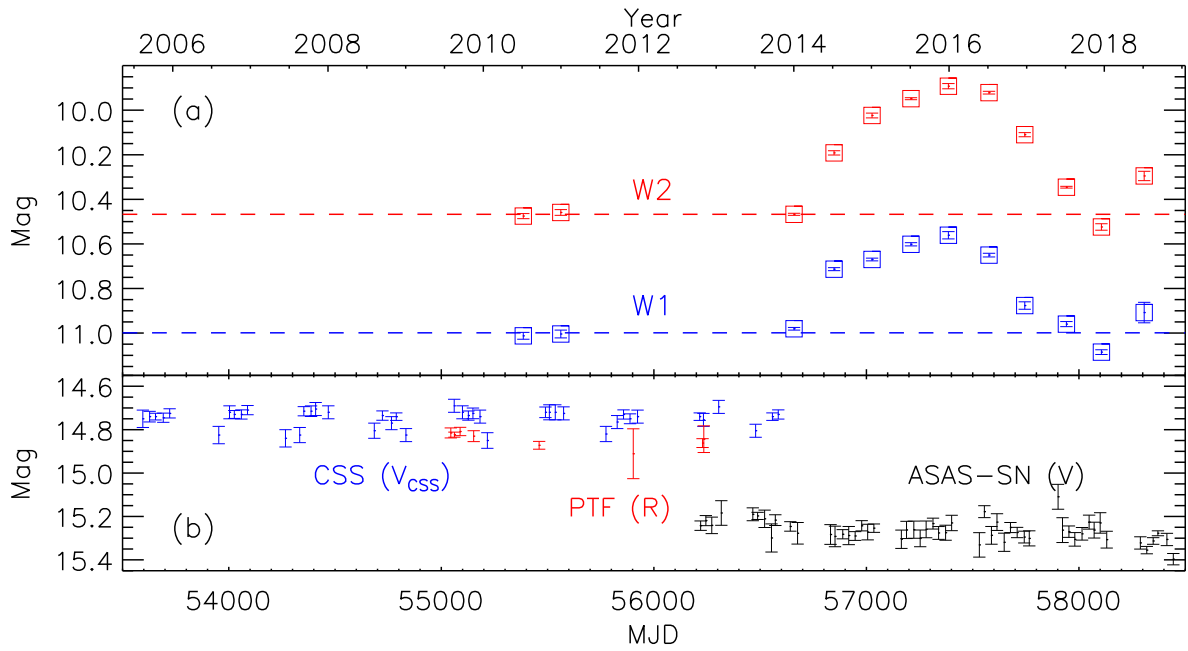}
 \caption{:
  {\bf (a)}: The MIR LCs of MCG-02-04-026 in W1 (red) and W2 (blue) bands.
  The LCs were constructed using PSF magnitudes.
  The horizontal dashed lines shows the levels in the two bands before the flare.
  {\bf (b)}: The binned optical LCs from surveys of CSS (blue), PTF (red) and ASAS-SN (black).
  }}
\label{fig1}
\end{figure}

\begin{figure}
\centering{
 \includegraphics[scale=0.8]{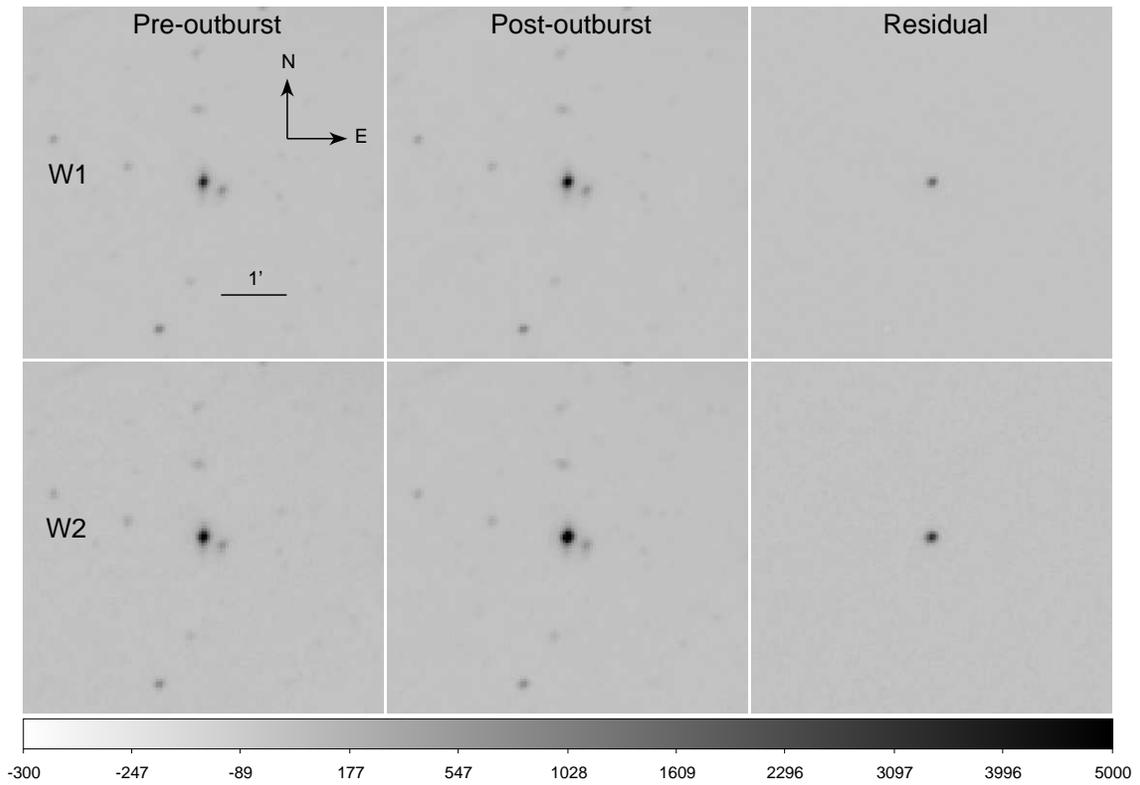}
 \caption{:
  The pre-flare images (left panels), flare images (middle panels), and the residual images (right panels) by subtracting the pre-flare images from the flare images.
  We show the images in W1 filter in the upper panels and images in W2 filter in the lower panels.
  The images are plotted in a uniform square-root scale, which is shown in the greyscale bar.
  }}
\label{fig2}
\end{figure}

\begin{figure}
\centering{
 \includegraphics[scale=0.98]{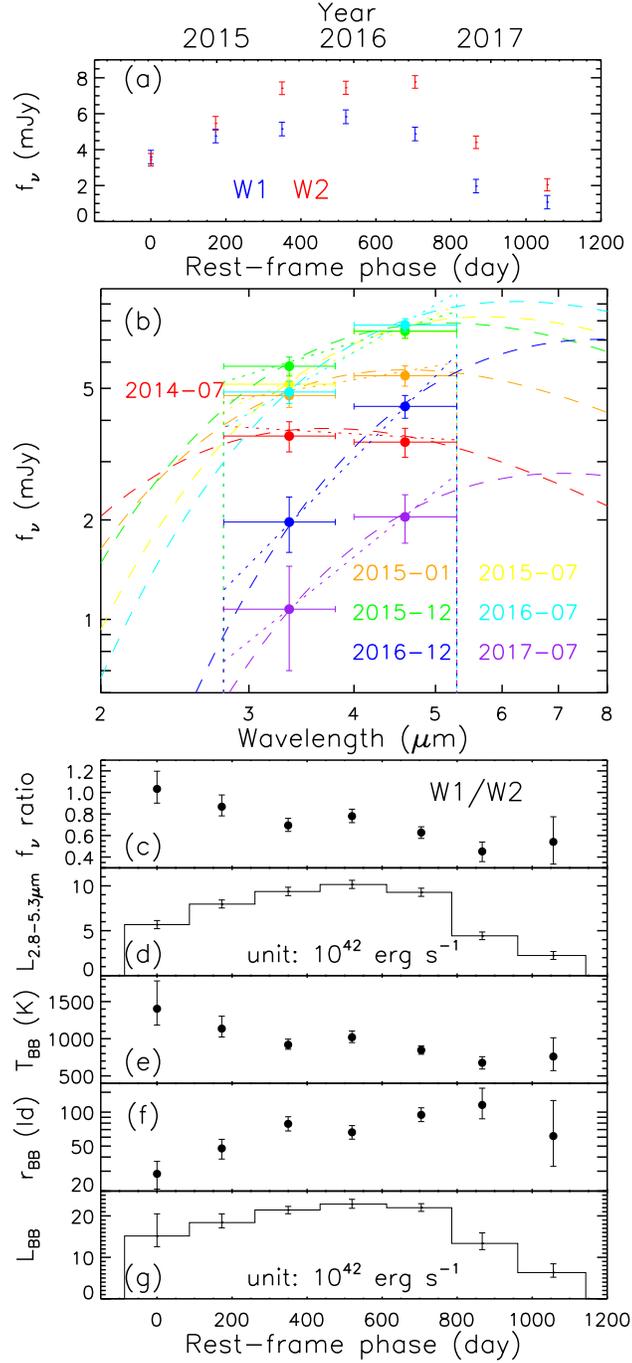}
 \caption{:
  {\bf (a)}: The LC of the MIR flare in W1 and W2 bands during July 2014 and July 2017.
  The phase was calculated relative to the time of the first excess (MJD$=$56848), and was converted to the rest-frame.
  {\bf (b)}: The spectra of the flare (data points), the power-law models (dotted lines) and the blackbody models (dashed lines).
  {\bf (c)}: The MIR color expressed as the ratio between fluxes in W1 and W2 bands.
  {\bf (d)}: The IR luminosity in the wavelength range of 2.8--5.3 $\mu$m.
  {\bf (e)(f)}: The blackbody temperature and the blackbody radius.
  {\bf (g)}: The luminosity calculated using blackbody models.
  }}
\label{fig3}
\end{figure}

\begin{figure}
\centering{
 \includegraphics[scale=0.82]{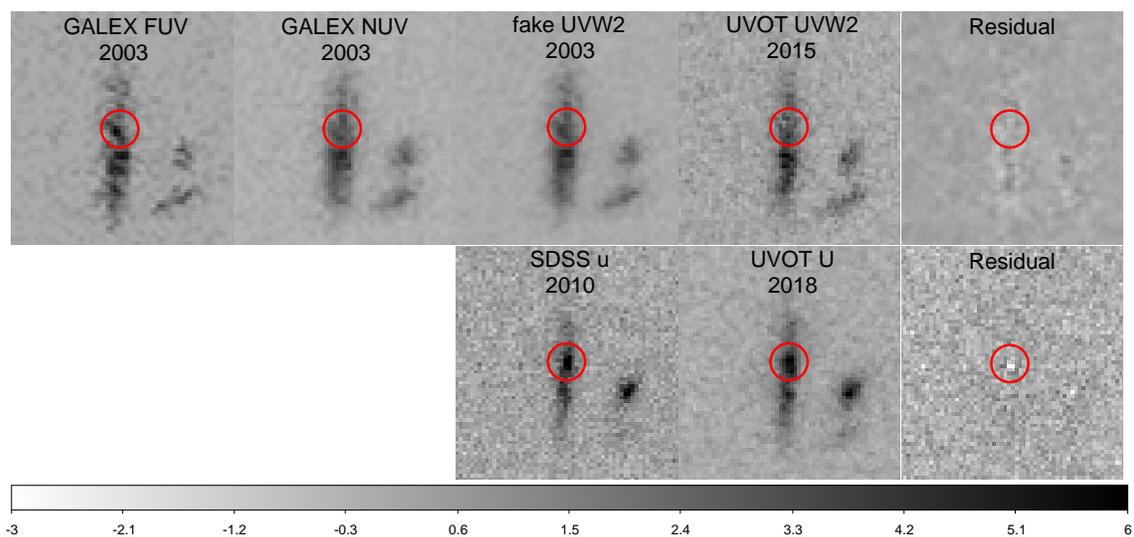}
 \caption{:
  We show the pre-2014, post-2014 and residual images in the two pairs of UV images.
  Those in the first pair are shown in the first row, and those in the second pair are in the second row.
  We show the pre-2014 images in the 3rd column, post-2014 images in the 4th column, and residual images in the 5th column.
  We also show the two GALEX images using which the pre-2014 image was obtained in the 1st and 2nd columns in the first row.
  For each image, we labeled which facility and filter were used in the observation, and the year of the observation.
  We also labeled the center of the galaxy using red circles with radii of 5$\arcsec$.
  All the images are normalized and are plotted in a uniform linear scale, which is shown in the greyscale bar.
  }}
\label{fig4}
\end{figure}

\begin{figure}
\centering{
 \includegraphics[scale=0.96]{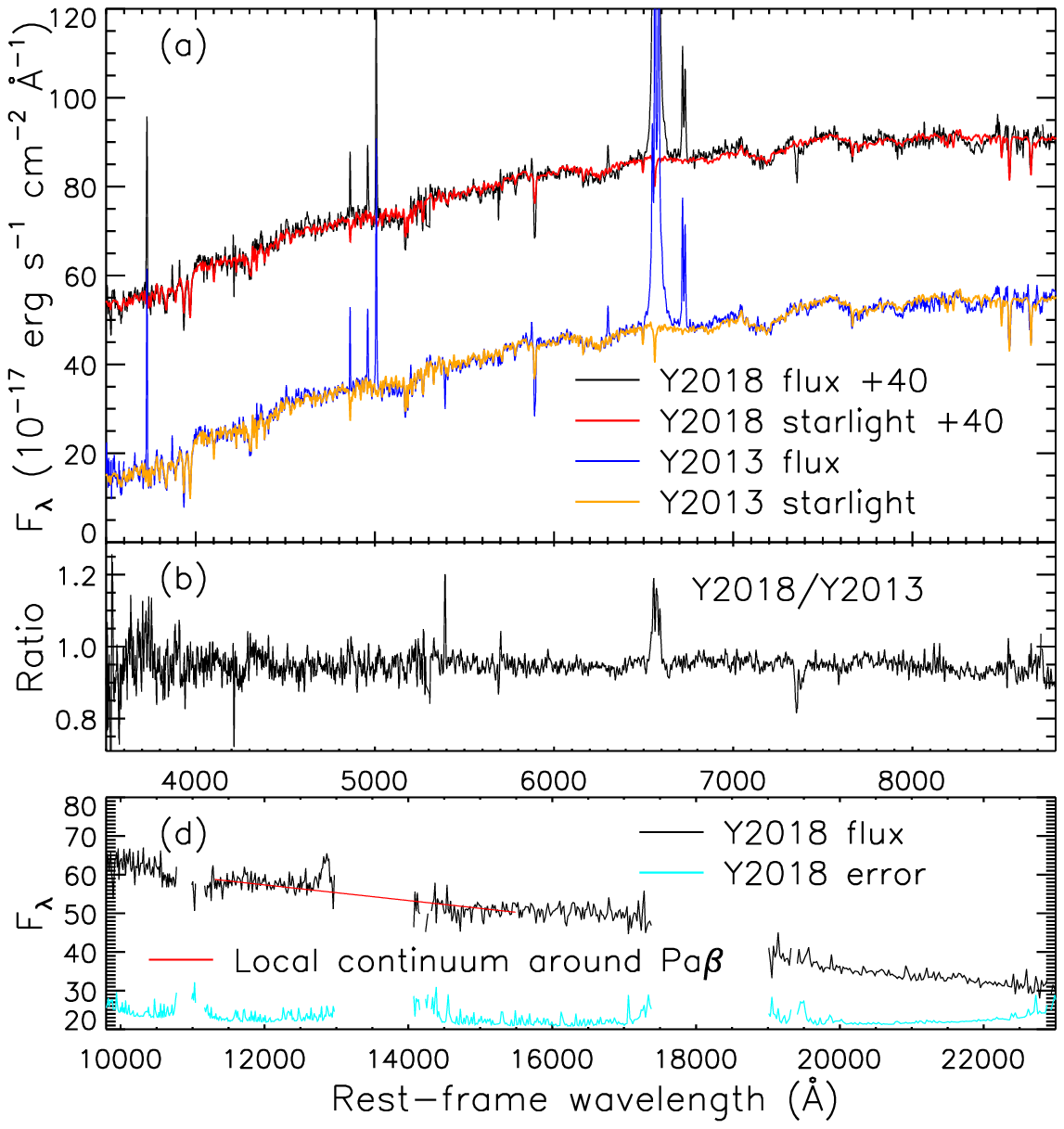}
 \includegraphics[scale=0.28]{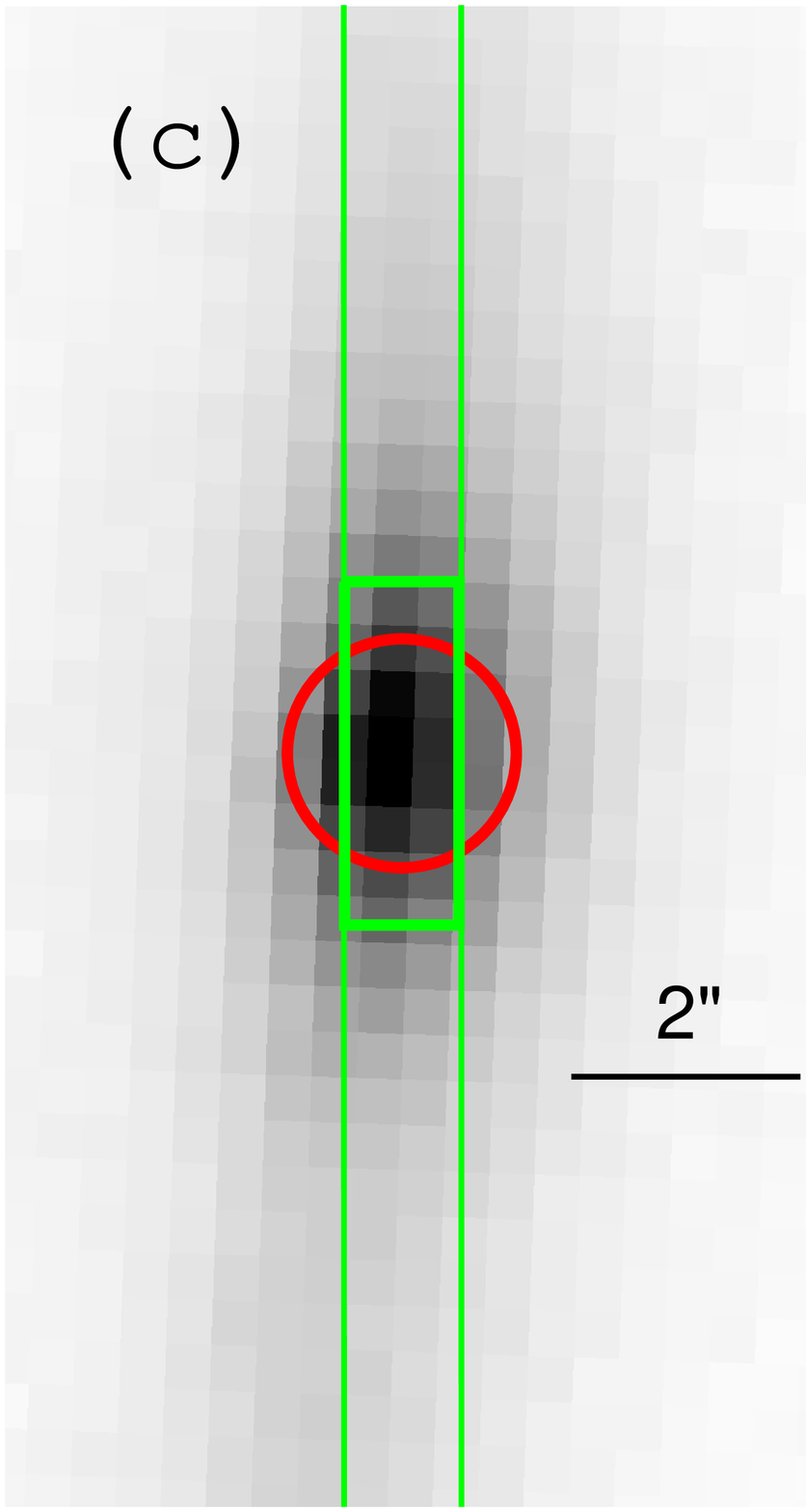}
 \caption{:
  {\bf (a)}: The optical spectra taken in 2013 (blue) and in 2018 (black), and the starlight models for them (orange for Y2013 and red for Y2018).
  {\bf (b)}: The ratio between Y2018 and Y2013 spectra.
  {\bf (c)}: The aperture (red circle) used in Y2013 observation with BOSS, and the slit and aperture (green lines) used in Y2018 observation with DBSP.
  {\bf (d)}: The NIR spectrum (black for flux and cyan for error) taken in 2018 with TPSP, and the local continuum model for region around Pa$\beta$ emission line.
  Here we displayed the binned spectrum in regions little affected by telluric absorptions for clarity.
  }}
\label{fig5}
\end{figure}

\begin{figure}
\centering{
 \includegraphics[scale=0.98]{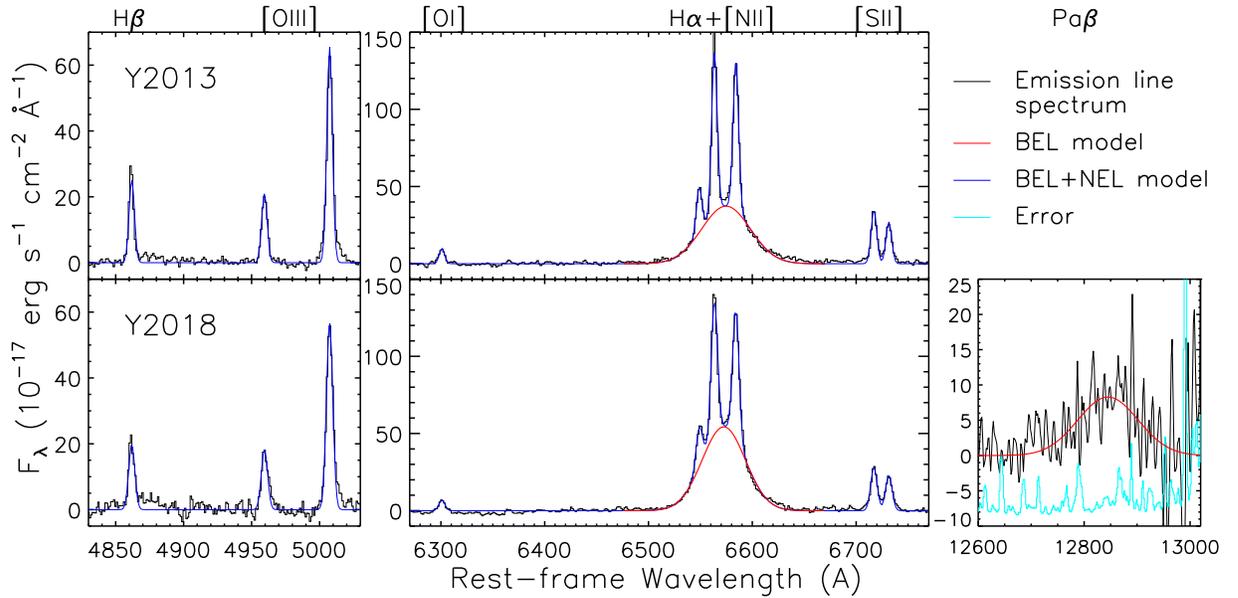}
 \caption{:
  The emission line spectra (black) from Y2013 (top panels) and Y2018 (bottom panels) observations and the models (blue), displayed in three wavelength ranges, the first is around H$\beta$+[OIII] (left panels), the second is around H$\alpha$ (middle panels), and the third is around Pa$\beta$ (right panel).
  The models contains BEL components and NEL components, and we show the BEL components in red lines.
  We also show the error of the NIR spectrum (minus 10) in cyan line.
  }}
\label{fig6}
\end{figure}

\begin{figure}
\centering{
 \includegraphics[scale=0.50]{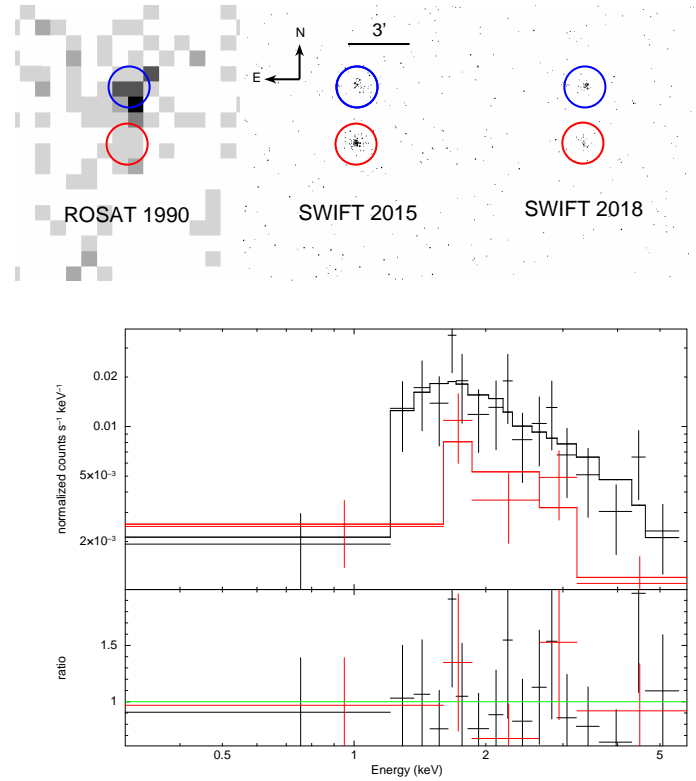}
 \includegraphics[scale=0.35,angle=270]{fig7b.ps}
 \caption{:
  {\bf Upper}: The X-ray images of MCG-02-04-026, one is from an observation taken in 1999 with ROSAT, and the other two are from observations taken in 2015 and 2018 with SWIFT.
  We label the positions of MCG-02-04-026 and the nearby Seyfert 1 galaxy using red and blue circles with radii of 1$\arcmin$.
  {\bf Lower}: X-ray spectra of MCG-02-04-026 from two observations in 2015 (black) and 2018 (red).
  For each spectrum, we plot the data (data point), and the simple absorbed power-law model (lines), and the ratio between the data and the model.
  }}
\label{fig7}
\end{figure}

\begin{figure}
\centering{
 \includegraphics[scale=0.98]{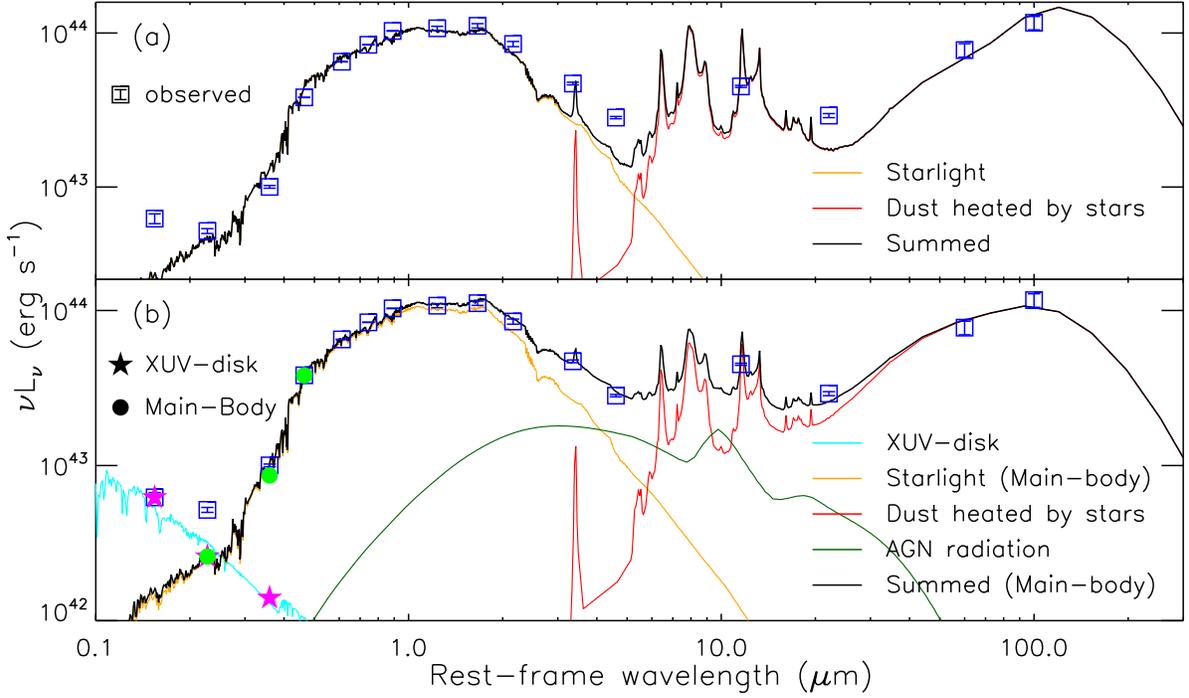}
 \caption{:
  {\bf (a)}: The pre-flare SED of MCG-02-04-026 (blue boxes and data points) and a simple fit to it.
  We fit it using a simple stellar radiation model (black line) which consists of a starlight component (orange) and a dust radiation component (red).
  {\bf (b)}: The final fit to the pre-flare SED.
  We decomposed the radiation of MCG-02-04-026 in the FUV to g bands into two components, a XUV-disk component (magenta pentagram) and a Main-body component (green circle).
  We show the simple stellar radiation model for the XUV-disk (cyan line).
  We also show the final model for the Main-body component (black line), which consists of a starlight component (orange), a stellar dust radiation component (red) and an AGN component (dark green).
  }}
\label{fig8}
\end{figure}

\begin{figure}
\centering{
 \includegraphics[scale=0.98]{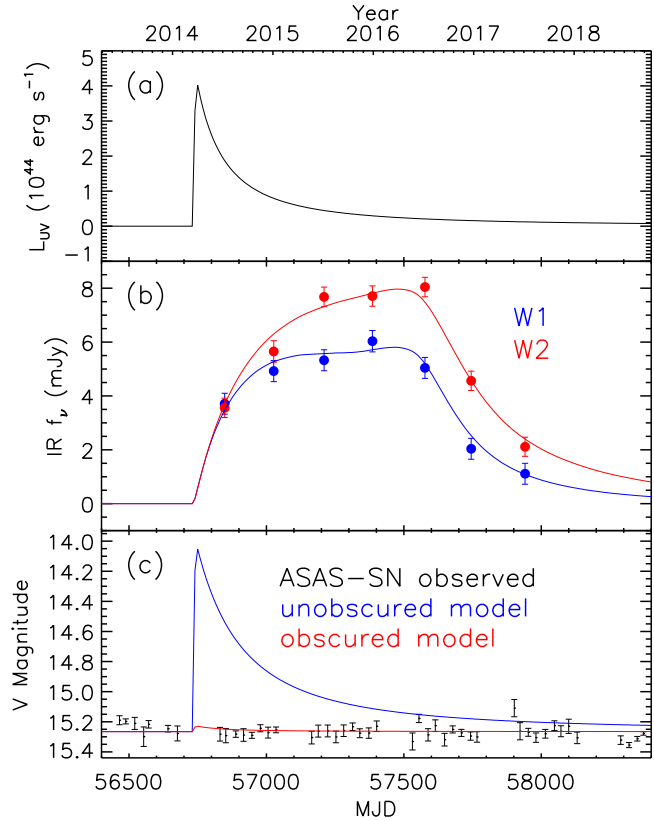}
 \caption{:
  The dust echo model interpreting the observations.
  {\bf (a)}: The predicted UV-optical LC of the primary transient events.
  {\bf (b)}: The predicted MIR LCs and the observed data.
  {\bf (c)}: The predicted optical V-band LC and the the LC from ASAS-SN observation.
  We show an unobscured model (blue) by assuming the UV-optical radiation has a spectral shape of a blackbody curve with $T=1.7\times10^4$ K, and an obscured model (red) by assuming a dust extinction of $A_V=4.7$ mag.
  }}
\label{fig9}
\end{figure}

\begin{figure}
\centering{
 \includegraphics[scale=0.98]{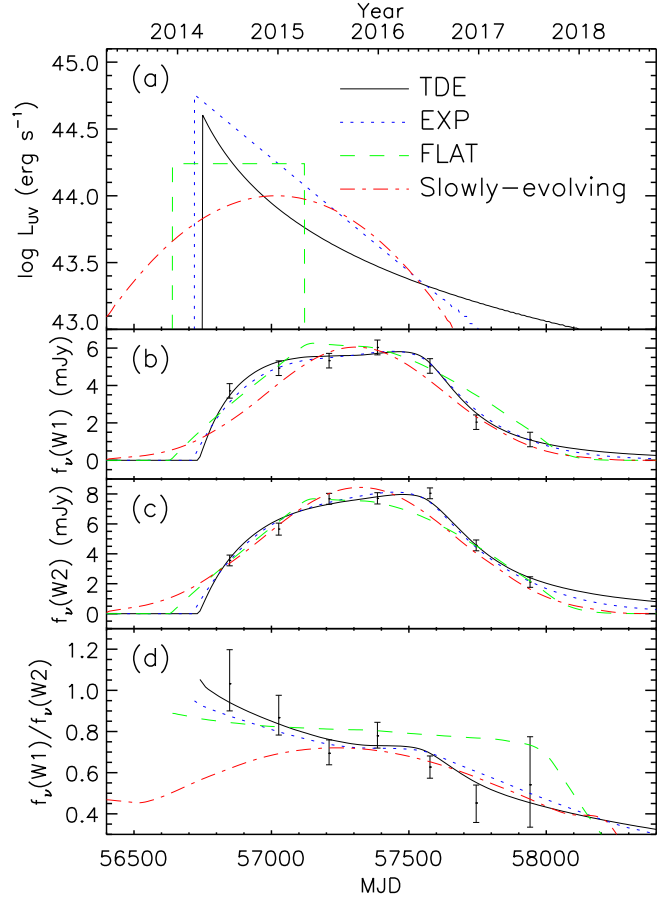}
 \caption{:
  The comparison of models by assuming different UV-optical LCs.
  We show those of TDE form in black solid lines, those of EXP form in blue dotted lines, those of FLAT form in green dashed lines, and those of showly-evolving form in red dash-dot lines.
  {\bf (a)}: The UV-optical LCs.
  {\bf (b)-(d)}: The predicted LC in W1 band, in W2 band, and the ratio between them, and the observation.
  }}
\label{fig10}
\end{figure}

\clearpage

\begin{appendix}

\setcounter{figure}{0}
\renewcommand{\thefigure}{A\arabic{figure}}

\section{The Approximate Empirical Formulas in the Dust Echo Model}

{ The calculation of the Lu16's model can be divided into 4 steps.
Step one, to calculate the variation of grains radius $a$ as functions of the distance $R$ and retarded time $t_r$.
Step two, to calculate the variation of the dust temperature $T$ and emissivity $j_{\rm IR}(\lambda)$ as functions of $R$ and $t_r$.
Step three, to calculate the variation of the attenuation of the IR radiation by other dust shells along the line of sight $\tau_{\rm IR}(\lambda)$ as functions of position ($R$, $\theta$) and $t_r$.
Step four, to integrate the attenuated IR radiation to obtain observed light curves.
Among these four steps, the first and third steps are the most time-consuming.
In the first step, one need to numerically solve the time-dependent differential equations.
In the third step, one need to track the IR photons and collect the properties of dust grains which the photons pass by (note that these properties vary with time) to calculate the attenuation.
Thus, we used empirical formula in the first and third steps.}

{ First we studied how the survival time of dust grains ($t_{\rm surv}$) varies with distance $R$.
We show the results from rigorous simulations in Figure A1.
We found that in a large space within the sublimation radius, $t_{\rm surv}$ grows nearly exponentially with the increase of $R$.
We use the following functions as approximate empirical formulas (red lines in Figure A1):}
\begin{equation}
t_{\rm surv}(R) =  \left\{
     \begin{array}{lr}
       { t_{\rm flare} e^{\alpha(\frac{R}{R_{\rm sub}}-1)},\ R_{\rm in}<R<R_{\rm sub} } \\
       { \infty,\ R>R_{\rm sub} }
     \end{array} \right.
\end{equation}
{ The parameter $t_{\rm flare}$ stands for the duration of the high state of the transient event, and is set to be:}
\begin{equation}
t_{\rm flare} =  \left\{
     \begin{array}{lr}
       { 0.15 E_{\rm tot}/L_{\rm max},\ {\rm TDE\ form} } \\
       { 0.3 E_{\rm tot}/L_{\rm max},\ {\rm EXP\ form} } \\
       { E_{\rm tot}/L_{\rm max},\ {\rm FLAT\ form} }
     \end{array} \right.
\end{equation}
{ The parameter $\alpha$ depends on the form of UV-optical LC: it is $\sim$16.5 for FLAT form and $\sim$19 for EXP and TDE form.
We fixed the $\alpha$ value according to the form of UV-optical LC.
We note that for $R<0.8R_{\rm sub}$, $t_{\rm surv}$ from the formula deviates from that from rigorous simulations.
However, this deviation has little effect on the final results because the dust grains at positions $R<0.8R_{\rm sub}$ are destroyed soon and radiate little IR energy.}


{ Then we studied how the grain radius ($a$) varies with time.
We show the results at four positions ($R/R_{\rm sub}=$0.8, 0.9, 0.99 and 1.1) from rigorous simulations in Figure A2.
For the first three positions, the time is normalized by the local grain survival time ($t_{\rm surv}(R)$), and for the last position, the time is normalized by the cutoff time of our simulation.
We found that the variation in the grain radius (a) in positions inside the $R_{\rm sub}$ show a consistent pattern: the grain radius first remains almost unchanged, and then gradually decreases with increasing rate, and finally rapidly decreases to zero.
We also found that the grain radius in positions outside the $R_{\rm sub}$ remains almost the same as the initial value.
We described the variation of $a$ with approximate empirical formulas.
For grains in positions with $R<R_{\rm sub}$, we used the following piecewise function (black line in Figure A2):}
\begin{equation}
a(R,t_r) =  \left\{
     \begin{array}{lr}
       { a_0,\ x<0.5 } \\
       { a_0(1.1-0.2x),\ 0.5<x<0.8 } \\
       { a_0(1.42-0.6x),\ 0.8<x<1 } \\
       { 0,\ x>1 }
     \end{array} \right.
\end{equation}
{ where $x=\frac{t_r}{t_{\rm surv}(R)}$.
And for grains with $R>R_{\rm sub}$, we set:}
\begin{equation}
a(R,t_r)=a_0
\end{equation}

{ Finally we studied how the IR optical depth $\tau_{\rm IR}(\lambda)$ varied with position ($R$, $\theta$) and retarded time ($t_r$).
After testing, we found that one can make assumptions described by the following formula:}
\begin{equation}
\tau_{\rm IR}(R,\theta,t_r,\lambda) =  \left\{
     \begin{array}{lr}
       { \tau_{\rm IR}(R,\theta,\lambda,t_r=0),\ \theta \leq 90^\circ } \\
       { \tau_{\rm IR}(R,\theta,\lambda,t_r=\infty),\ \theta > 90^\circ }
     \end{array} \right.
\end{equation}
{ The meaning of the formula is: to always use the $\tau_{\rm IR}$ value in the initial state for the hemisphere close to us, and to always use the $\tau_{\rm IR}$ value in the final stable state for the hemisphere away from us.
Our test shows that making these assumptions has little effect on the final results.}

{ We verified the validity of our model by comparing the LCs and SEDs of IR radiation from the rigorous simulations, and those from simulations in which the above approximate empirical formulas used.
We found that the differences are within 2\% in most cases, as shown by an example in Figure A3, and within 5\% in extreme cases.}

\begin{figure}
\centering{
 \includegraphics[scale=0.98]{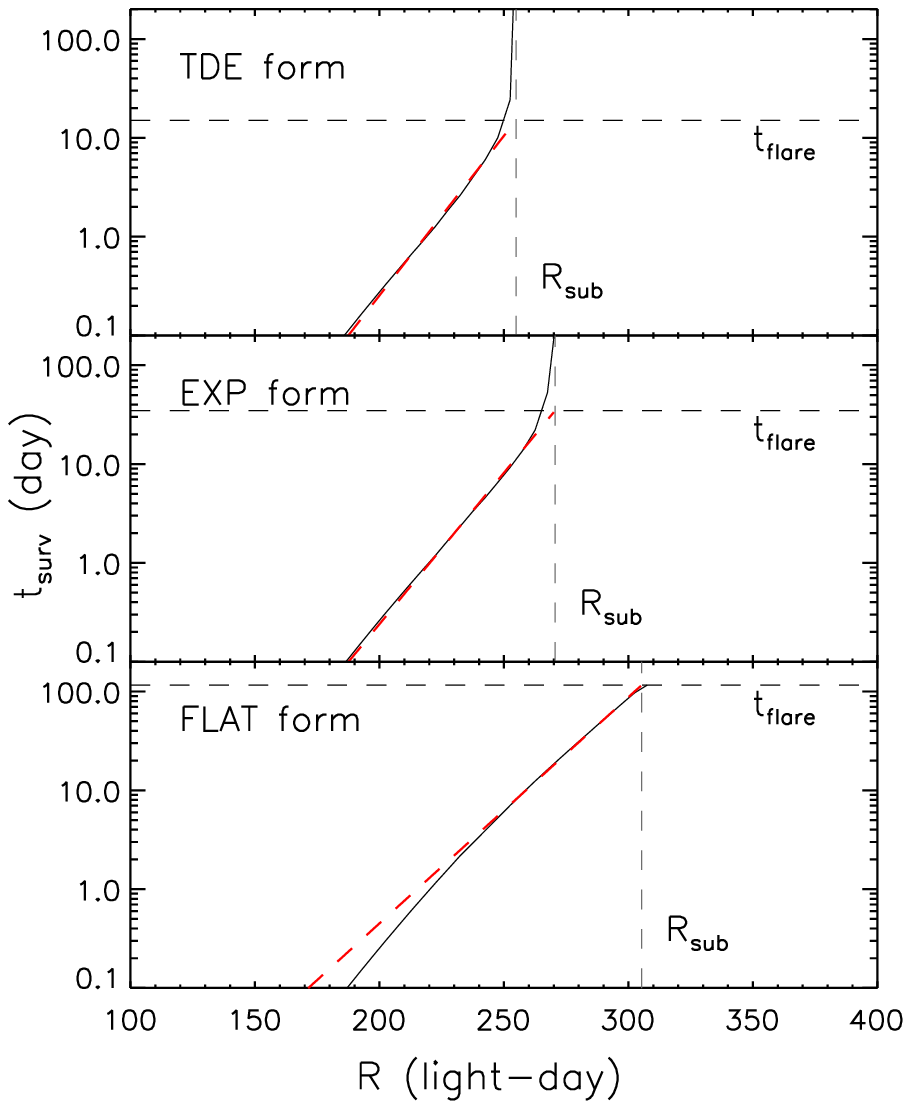}
 \caption{:
  The survival time of dust grains ($t_{\rm surv}$) varies with distance $R$.
  We show the results in three situations.
  The three situations assumed TDE, EXP and FLAT forms UV-optical LCs, respectively, and in all of the three situations, $L_{\rm max}=10^{45}$ erg s$^{-1}$, $E_{\rm tot}=10^{52}$ erg, $R_{\rm in}=100$ light-days, $R_{\rm out}=600$ light-day, $a_0=0.1$ $\mu$m, $n_d=10^{-8}$ cm$^{-3}$.
  We show the results from rigorous simulations with black solid lines, and the approximate empirical formulas with red lines.
  }}
\label{figA1}
\end{figure}

\begin{figure}
\centering{
 \includegraphics[scale=0.98]{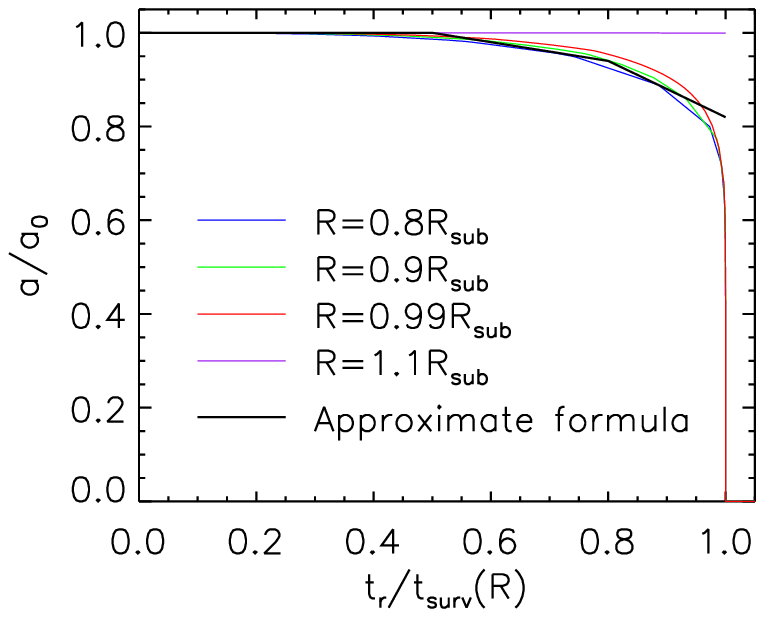}
 \caption{:
  The grain radii (a) varies with retarded time ($t_r$).
  We show the results for dust grains in positions with distances $R$ of 0.8, 0.9, 0.99 and 1.1 $R_{\rm sub}$ (blue, green, red and purple lines) in the third situation described in the caption of Figure A1.
  We also show the piecewise function we defined as the approximate empirical formula.
  }}
\label{figA2}
\end{figure}

\begin{figure}
\centering{
 \includegraphics[scale=0.98]{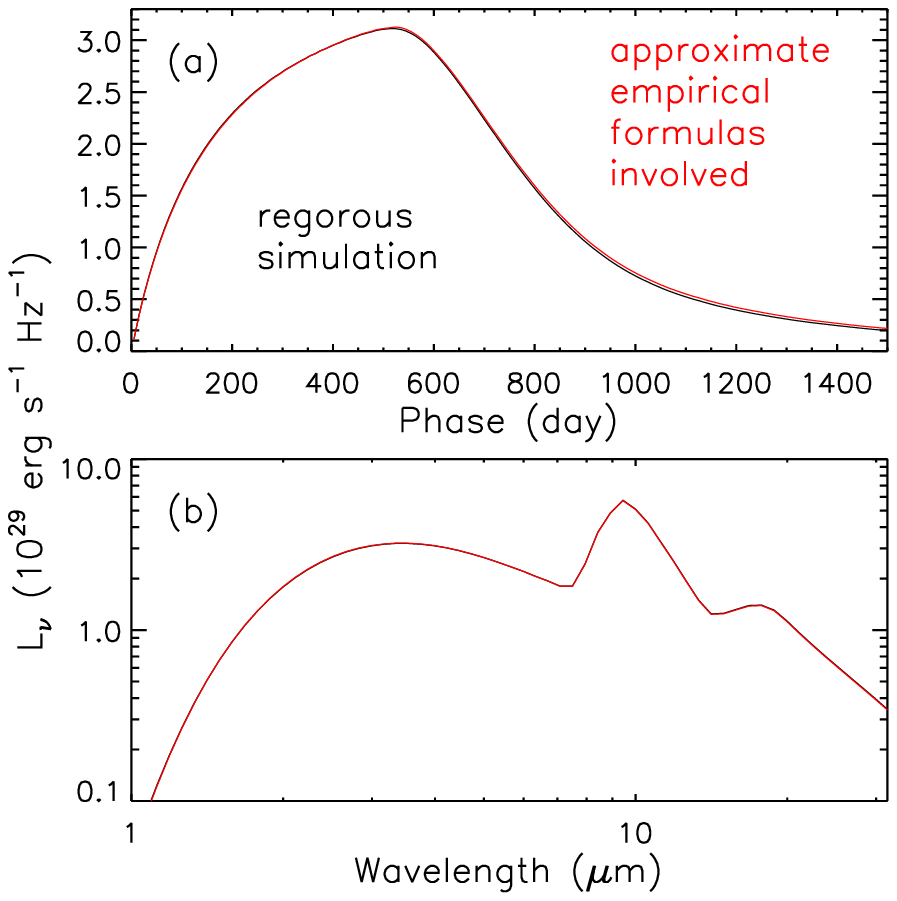}
 \caption{:
  The comparison between the results of the rigorous simulations (black lines) and simulations with the approximate empirical formulas involved (red lines).
  We show the results of simulation in the first situation described in the caption of Figure A1.
  We show the LCs at 4$\mu$m in the upper panel, and SEDs on the 510th days (close to the peak) in the lower panel.
  }}
\label{figA3}
\end{figure}

\end{appendix}

\end{document}